\documentclass[twocolumn,aps,longbibliography]{revtex4-2}
\usepackage{dcolumn,graphicx,bm,hyperref,amsmath,amssymb,xcolor,booktabs,caption,romannum,float,subcaption,diagbox,array} 
\usepackage{natbib,bm}
\usepackage[ruled,vlined]{algorithm2e}
\usepackage{tikz}
\usepackage{qcircuit,braket}

\begin{document}  

\title{Subsampling Factorization Machine Annealing  }  

\author{Yusuke Hama}
 \affiliation{Global R$\&$D Center for Business by Quantum-AI Technology (G-QuAT), National Institute of Advanced Industrial Science and Technology (AIST), 1-1-1 Umezono, Tsukuba, Ibaraki 305-8568, Japan}%
 \email{hama.yusuke@aist.go.jp}
\author{Tadashi Kadowaki}
\affiliation{DENSO CORPORATION, 1-1-4, Haneda Airport, Ota-ku, Tokyo 144-0041, Japan}%
 \affiliation{Global R$\&$D Center for Business by Quantum-AI Technology (G-QuAT), National Institute of Advanced Industrial Science and Technology (AIST), 1-1-1 Umezono, Tsukuba, Ibaraki 305-8568, Japan}%
\begin{abstract}
{Quantum computing and machine learning are state-of-the-art technologies that have been investigated intensively in both academia and industry.  
The hybrid technology of these two ingredients is expected to be a powerful tool to solve complex problems in many branches of science and engineering such as combinatorial optimization problems and accelerate the creation of  next-generation technologies. 
 In this work, we develop an algorithm to solve a black-box optimization problem by improving Factorization Machine Annealing (FMA) such that  the training of a machine learning model called Factorization Machine is performed not by a full dataset but by a subdataset  that is sampled from a full dataset: Subsampling Factorization Machine Annealing (SFMA). According to such a probabilistic training process, 
the performance of FMA on exploring a solution space gets enhanced.  As a result,    SFMA exhibits  balanced performance of exploration and exploitation, which we call exploitation-exploration functionality.   
We conduct numerical benchmarking tests to compare the performance of  SFMA  with that of  FMA. Consequently,  SFMA certainly exhibits the exploration-exploitation functionality and outperforms FMA in speed and  accuracy. 
In addition, the performance of SFMA can be further improved by sequentially using two subsampling datasets with different sizes such that the size of the latter dataset is substantially smaller than the former.  Such a substantial reduction not only enhances the exploration performance of SFMA but also enables us to run it with correspondingly low computational cost even for a large-scale problem. 
These  results indicate the effectiveness of  SFMA in a certain class of black-box optimization problems  of significant  size: the potential scalability of SFMA in solving large-scale problems with correspondingly low computational cost. 
     The exploitation-exploration functionality of SFMA is expected to pave the way for solving various complex combinatorial optimization problems  in the real world and designing next-generation technologies that become the basic building blocks for industrial advancements.  }
\end{abstract}
  
 \maketitle

\section{\label{intro}Introduction} 
Optimization problems are ubiquitous in science and engineering (physics, chemistry, computer science, etc.) and have been studied extensively for long decades \cite{mezard2009information,kochenderfer2019algorithms,lucas2014ising,suh2022convex,yarkoni2022quantum,phillipson2024quantum}.
It includes various kinds of fundamental problems such as traveling salesperson problem, max-cut problem, and graph coloring problem, 
and the algorithms for solving these problems are expected to be the basic building blocks for advancing industrial technologies such as logistics and production planning  \cite{yarkoni2022quantum,phillipson2024quantum}.

In physics, combinatorial optimization problem has been investigated intensively by using a powerful and simple formalism called quadratic unconstrained binary optimization (QUBO), or equivalently, Ising model \cite{mezard2009information,lucas2014ising,yarkoni2022quantum,phillipson2024quantum,das2008colloquium,tanaka2017quantum,albash2018adiabatic,hauke2020perspectives,rajak2023quantum}.
The canonical algorithms for the optimization based on QUBO (Ising model)  include simulated annealing (SA),  quantum annealing (QA), and/or adiabatic  quantum computing \cite{kochenderfer2019algorithms,lucas2014ising,yarkoni2022quantum,phillipson2024quantum,das2008colloquium,tanaka2017quantum,wendin2017quantum,albash2018adiabatic,hauke2020perspectives,rajak2023quantum,
kirkpatrick1983optimization,bertsimas1993simulated,kadowaki1998quantum,brooke1999quantum,farhi2001quantum,aharonov2008adiabatic,johnson2011quantum,boixo2014evidence,ronnow2014defining,
boixo2016computational,denchev2016computational,albash2018demonstration,tsukamoto2017accelerator,aramon2019physics,au2024quantum,king2025beyond}. On the other hand, in gate-based quantum computing variational algorithms so-called quantum approximate optimization algorithm and quantum alternating operator ansatz have been developed and utilized for solving combinatorial optimization problems \cite{au2024quantum,farhi2014quantum,moll2018quantum,hadfield2019quantum,sachdeva2024quantum,blekos2024review,abbas2024challenges}.
Furthermore, owing to the advancement of both classical and quantum computers (quantum annealer and gate-based quantum computer), such physics techniques become able to be implemented on these machines and solve many types of optimization problems related to societal issues as well as the development of  technologies, for instance, materials engineering, drug discovery, portfolio optimization, and protein folding  \cite{yarkoni2022quantum,wendin2017quantum,moll2018quantum,camino2023quantum,hatakeyama2022automated,hatakeyama2023extracting,gao2023quantum,sampei2023quantum,mcardle2020quantum,bauer2020quantum,lee2023evaluating,
venturelli2019reverse,lang2022strategic,buonaiuto2023best,aguilera2024multi,
perdomo2008construction,perdomo2012finding,babbush2014construction,irback2022folding,irback2024using}.    
Meanwhile, the development of machine learning techniques or algorithms (both classical and quantum) has been progressing \cite{suh2022convex,bishop2006pattern,bengio2017deep,mehta2019high,murphy2022probabilistic,tian2023recent,bishop2023deep,
hopfield1982neural,ackley1985learning,hinton2002training,amin2018quantum,date2021qubo,biamonte2017quantum,dunjko2018machine,benedetti2019parameterized,schuld2021machine,cerezo2022challenges,sajjan2022quantum,zeguendry2023quantum,wang2024comprehensive}, and such data-driven approaches have been playing significant roles in, for instance, simulations and designing in materials science and engineering, i.e., materials informatics (MI) \cite{agrawal2016perspective,ramprasad2017machine,alberi20182019,agrawal2019deep,hong2021reducing,iwasaki2021machine,merchant2023scaling,li2023methods,chen2024accelerating,sivan2024advances}.   
Due to such  advancements,  in recent years the hybrid techniques of quantum computing (quantum annealing) and machine learning have been developed, for example, so-called Factorization Machine  Annealing (FMA) \cite{kitai2020designing,seki2022black,wilson2021machine,inoue2022towards,matsumori2022application,kadowaki2022lossy,mao2023chemical,tucs2023quantum,nawa2023quantum,liu2024implementation,couzinie2025machine,lin2025determination,liang2025crysim,nakano2026swift}, variants of FMA  \cite{minamoto2025black,kashimata2025real}, and
Bayesian Optimization of Combinatorial Structures (BOCS) \cite{baptista2018bayesian,koshikawa2021benchmark,matsumori2022application,kadowaki2022lossy,morita2023random}.  
These  methods have been applied to solve a problem called black-box optimization (BBO), which is an  optimization problem where a functional form of an objective function is unknown. 
In most actual cases, the  optimization problems are considered to be BBO since the objective functions (output variables) are generated from  input variables through complex processes. 
For instance, the optimization of  physical properties of target composite materials is BBO in general, because how the physical properties (objective functions)  are generated from  ingredients (input variables) are complex processes \cite{kitai2020designing,iwasaki2021machine,nawa2023quantum,lin2025determination,sampei2023quantum}. Both FMA and BOCS have been applied to materials simulations and designing and many successful results have been reported  \cite{kitai2020designing,seki2022black,wilson2021machine,inoue2022towards,matsumori2022application,mao2023chemical,tucs2023quantum,nawa2023quantum,couzinie2025machine,lin2025determination,liang2025crysim}.
Since the advancement of the quantum computing and machine learning techniques is expected to continue, 
it is an important issue to further develop  effective hybrid techniques or algorithms of these two ingredients. 
It is expected that such hybrid algorithms are going to be powerful tools to tackle a wide range of optimization problems in the real world. 
 
Toward the above goal, in this work we develop an effective algorithm for BBO by improving FMA.
In general, the key to effectively solve combinatorial optimization problems  is to coherently perform exploration and exploitation in  solution spaces.
That is,  we search candidate solutions in a wide range of solution spaces in the first half of  optimization processes  (exploration), while we concentrate on finding the best candidate solutions in the second half (exploitation) \cite{kochenderfer2019algorithms,kim2023quantum}.
On the other hand,  we consider that FMA has  good performance on exploitation but not  on exploration since it is basically a point-estimation approach with regard to machine learning parameters. 
By taking this limitation into account, we improve FMA by amplifying the performance of exploration,
which is done by training a machine learning model with a subdataset that is sampled from a full dataset. 
We call this algorithm Subsampling Factorization Machine Annealing (SFMA). 
In SFMA, a machine learning model is trained probabilistically and  the exploration performance of FMA gets enhanced. 
As a result,  SFMA  exhibits the balanced performance of exploration and exploitation: exploration-exploitation functionality.  
To  numerically check whether SFMA truly possess  the exploration-exploitation functionality  and test its utility in numerous black-optimization problems,
we perform numerical experiments on  black-box optimization problems called lossy compression of data matrices \cite{kadowaki2022lossy,ambai2014spade,yoon2022lossy}.
These experiments are conducted by benchmarking SFMA against FMA.
Consequently,  SFMA certainly possesses the exploration-exploitation functionality and exhibits the faster convergence to the optimal solutions with higher accuracy than FMA.
Moreover, its exploration performance can be further enhanced by serially utilizing two different subsampling datasets such that the size of the latter dataset is significantly smaller than the former. 
 Namely, we are able to solve a large-scale problem by tailoring the size of a subdataset to be correspondingly small, which implies that SFMA can be run with correspondingly low computational cost. 
 As a result, the exploitation performance (accuracy) of SFMA is amplified as well. 
 This is the significant finding in our results and is expected to be a powerful tool for solving a complex problem in the real world. 
   
The paper is structured as follows.  In Sec. \ref{BBOFMA}, we briefly summarize the concept of BBO and the framework of FMA. 
In Sec.  \ref{SFMA}, which is one of the main sections of this paper,  we explain how to develop SFMA.  Furthermore, we discuss the advantages of SFMA with respect to 
computational cost and scalability.
In Sec. \ref{numexps}, which is the second main section of this paper, we demonstrate the numerical experiments on SFMA and analyze the results. 
Section \ref{conclusion} is devoted to the conclusions and outlook of this paper.

 \section{\label{BBOFMA} Black-Box Optimization and Factorization Machine Annealing} 
 
To understand how  Subsampling Factorization Machine Annealing (SFMA) is developed (described in detail  in Sec. \ref{SFMA}), in this section,
first we explain the concept of black-box optimization.  
After that, we explain the framework of FMA and its characteristic from a viewpoint of exploration and exploitation.   
 In the following, we assume that the optimal solution to a black-box function is the global minimum.

\begin{figure}
\centering
\includegraphics[width=8.5cm]{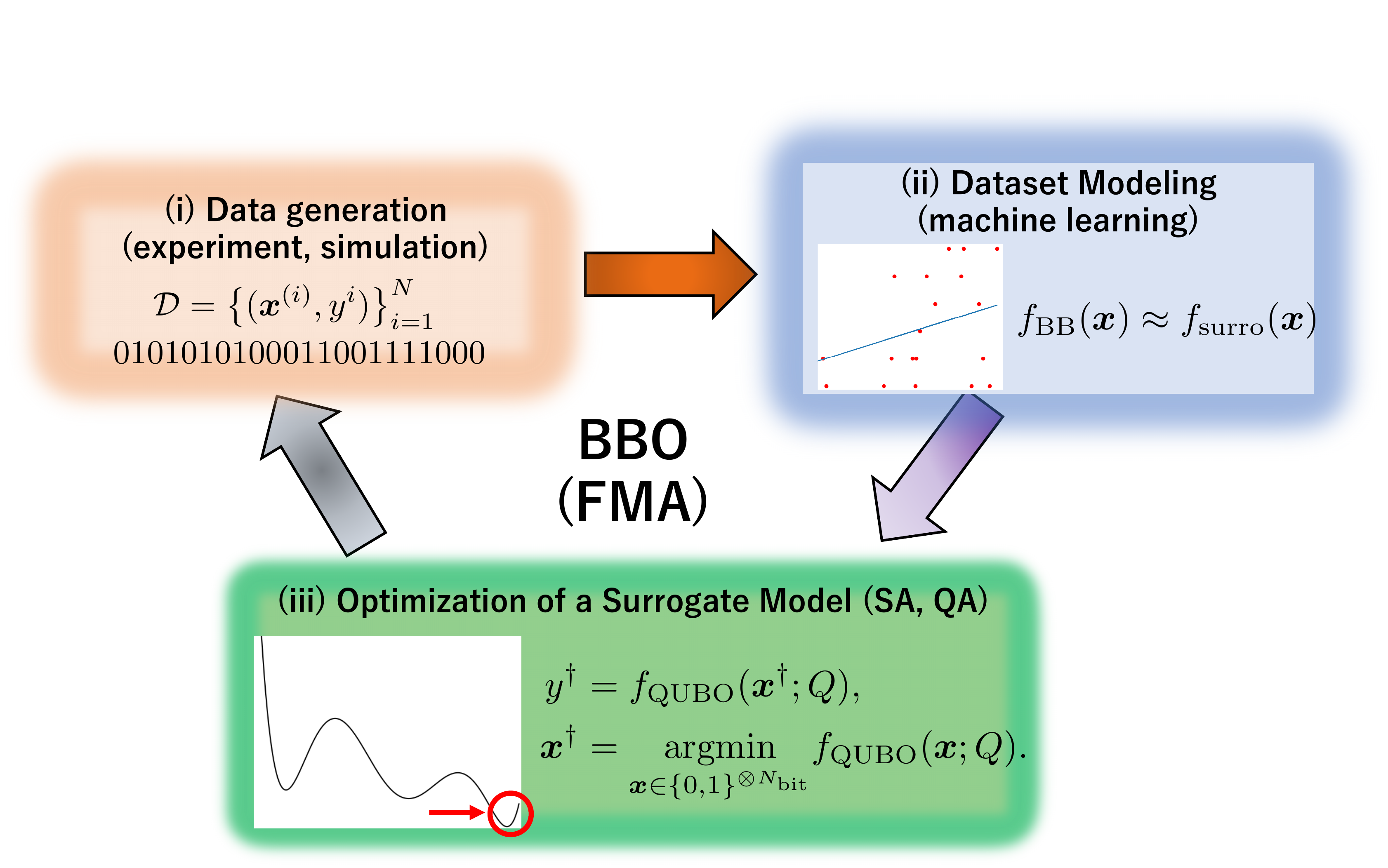}
\caption{Schematic of a procedure to conduct BBO. It is comprised of three steps: (i) data generation, (ii) dataset modeling, and (iii) optimization of a surrogate model.} 
\label{BBOschematic}
\end{figure}
\subsection{\label{BBO}Black-Box Optimization} 
The goal of BBO is to find the optimal solution to a black-box objective function that we denote by $f_\text{BB}$. 
Such an optimization problem is solved using a data-driven approach represented by a schematic in Fig. \ref{BBOschematic}.  BBO is conducted in three basic steps:  (i) data generation, (ii) dataset modeling, and (iii) optimization of a surrogate model. Step (i) is performed by, for instance, experiments or numerical simulations. Step (ii) is equivalent to the construction of a machine learning model or a surrogate model using a dataset at hand.  These three steps are performed iteratively. Hereinafter, let us call this sequential process BBO loop. In mathematical terms, suppose we have given $N_0$ data samples consisting of input (explanatory)  variables $\boldsymbol{x}^{(i)}$ and output (objective) variables $y^i (=f_\text{BB}(\boldsymbol{x}^{(i)}))$  with $i=1,\ldots,N_0$, i.e., the generation of an initial dataset in step (i). We describe such a dataset by a form $\mathcal{D}_{0}=\big{\{} (\boldsymbol{x}^{(i)},y^{(i)}) \big{\}}_{i=1}^{N_0}=\big{\{} (\boldsymbol{x}^{(1)},y^1), \ldots, (\boldsymbol{x}^{(i)},y^{(i)}), \ldots, (\boldsymbol{x}^{(N_0)},y^{(N_0)}) \big{\}}$. We assume $\boldsymbol{x}^{(i)}$ to be $N_\text{bit}$-component binary vectors, i.e., $\boldsymbol{x}^{(i)} = \big{(} x^{(i)}_1,  \ldots,  x^{(i)}_l,  \ldots,  x^{(i)}_{N_\text{bit}}       \big{)} $, where $x^{(i)}_l  \in \{0,1\}.$  
Next, in step (ii) we construct a surrogate model that we denote by $f_\text{surrogate}(\boldsymbol{x};Q^{(0)})$, where $Q^{(0)}$ denotes a collection of trainable parameters obtained by $\mathcal{D}_{0}$ and a machine learning algorithm. Then, in step (iii) we find the binary optimal solution to $f_\text{surrogate}(\boldsymbol{x};Q^{(0)})$ and denote it by $\boldsymbol{x}^{\dagger(1)}.$ After that, we add  $ \big{(} \boldsymbol{x}^{\dagger(1)}, y^{\dagger(1)} \big{)}$ to $\mathcal{D}_{0}$, where  
$y^{\dagger(1)}= f_\text{BB}(\boldsymbol{x}^{\dagger(1)})$. The dataset at hand is augmented such that $\mathcal{D}_{1} =  \big{\{} (\boldsymbol{x}^{(1)},y^{(1)}), \ldots,  (\boldsymbol{x}^{(N_0)},y^{(N_0)}), (\boldsymbol{x}^{(N_0+1)},y^{(N_0+1)}) \big{\}},$  where $\boldsymbol{x}^{(N_0+1)}  \equiv \boldsymbol{x}^{\dagger(1)}, y^{(N_0+1)}  \equiv y^{\dagger(1)}.$ We repeat the BBO loops  $N_\text{ite}-1$ ($N_\text{ite} \geq 2$) more times.  
As a result, we obtain the final dataset  $\mathcal{D}_{N_\text{ite}} =  \big{\{} (\boldsymbol{x}^{(1)},y^{(1)}), \ldots,  (\boldsymbol{x}^{(N_\text{tot})},y^{(N_\text{tot})}) \big{\}}$, where $N_\text{tot} = N_0 + N_\text{ite}$. Then,  we obtain an approximate optimal solution to $f_\text{BB}(\boldsymbol{x})$ that we denote by  
$(\tilde{\boldsymbol{x}}^{\ast},  \tilde{y}^{\ast})$, where $ \tilde{y}^{\ast}  = \text{min} \{ y^{(1)}, \ldots, y^{(N_\text{tot})} \} $.  
  In the following, we denote the true optimal solution by $(\boldsymbol{x}^{\ast},  y^{\ast})$ to distinguish it from $(\tilde{\boldsymbol{x}}^{\ast},  \tilde{y}^{\ast})$.

 Let us end this section by making a comment to avoid misunderstanding of BBO. In general, the goal of machine learning (supervised learning) is to train a model with a given dataset and use it  to make a quantitative prediction for new input data. To achieve this goal, machine learning is processed not only with training data but also with validation data and test data in order to avoid overfitting. In contrast, in BBO we use a trained (surrogate) model not for a prediction task  but for finding the global minimum of a black-box function under consideration. Therefore, we do not split the data  $\mathcal{D}_{a} $ into the training  and validation data to calculate a surrogate model. 
Instead,  the dataset at hand is fully used as training data and a surrogate model is built with it.   
\subsection{\label{FMA} Factorization Machine Annealing}
  FMA is the BBO algorithm where the surrogate model is Factorization Machine (FM) and the optimizer applied in step (iii) is an annealer such as SA and QA.
By denoting  $\boldsymbol{x}$ to be an $N_\text{ite}$-dimensional binary vector, the functional form of FM is given in terms of $\boldsymbol{x}$  by \cite{kitai2020designing,seki2022black,wilson2021machine,inoue2022towards,matsumori2022application,kadowaki2022lossy,mao2023chemical,tucs2023quantum,nawa2023quantum,liu2024implementation,couzinie2025machine,lin2025determination,liang2025crysim,nakano2026swift,rendle2010factorization}
\begin{footnotesize}
\begin{align}
f_\text{FM}(\boldsymbol{x}; \boldsymbol{\theta})= w_0 + \sum_{i=1}^{N_\text{bit}} w_i x_i  +   \sum_{  1 \leq i < j  \leq N_\text{bit}}  \left( \sum_{l=1}^k v_{i,l} v_{j,l}\right) x_i x_j, 
\label{FMformula1}
\end{align} \end{footnotesize}
where  $x_i$ ($x_j$) is the $i$th ($j$th) component  of $\boldsymbol{x}$,  $\boldsymbol{\theta} = (w_0,\ldots, w_{N_\text{bit}}, v_{1,1}, \ldots,  v_{N_\text{bit},k})^T$ are  trainable parameters of FM (FM parameters) that are real,  and  $k$ is a hyperparameter that represents the expressivity of  FM. 
By using the relation $x_i^2 =x_i $, we see that $f_\text{FM}(\boldsymbol{x}; \boldsymbol{\theta})$ in Eq. \eqref{FMformula1} has the same mathematical structure with a QUBO function. That is, by  writing  matrix elements of QUBO by $Q_{i,j} $, we observe that $Q_{i,i}=w_i $ and $Q_{i,j}= \sum_{l=1}^k v_{i,l} v_{j,l}$  ($i \neq j$). Although  $f_\text{FM}(\boldsymbol{x}; \boldsymbol{\theta})$ in Eq. \eqref{FMformula1} includes the constant term $w_0$, this does not affect optimization processes and annealers  can be utilized.     
 In FMA, the computational complexity to learn the coefficients $v_{i,l}$ (off-diagonal matrix elements of FM) is $\mathcal{O}(N_\text{bit}k)$, while the number of off-diagonal elements of  a QUBO matrix is $N_\text{bit}(N_\text{bit}-1)/2$, i.e., the computational cost to train off-diagonal elements of QUBO matrices is $\mathcal{O}(N_\text{bit}^2.)$  By tuning the hyperparameter $k$ to be $N_\text{bit}/2-1$ or less than that, we become able to 
 train FM functions with lower computational complexity than that for training general QUBO functions. 
 This is one of the advantages of adopting FM as a surrogate model. 

 Let us end this section by discussing the performance of FMA from a viewpoint of the exploration and exploitation performance. 
 FM models $f_\text{FM}(\boldsymbol{x}; \boldsymbol{\theta})$  (FM parameters $\boldsymbol{\theta}$) are trained by minimizing a loss function [e.g., mean square error (MSE)] using a gradient-based optimization method such as stochastic gradient descent (SGD) and Adam. In such a scheme, $f_\text{FM}(\boldsymbol{x}; \boldsymbol{\theta})$   are trained deterministically for a definite dataset: From this perspective, it can be considered that FM parameters are trained by a point-estimation approach. 
 Such a way of training surrogate models is in contrast to the approach taken in BOCS. That is, in BOCS, the matrix elements of QUBO matrices (surrogate-model parameters) are sampled from  a posterior probability distribution \cite{baptista2018bayesian,koshikawa2021benchmark,matsumori2022application,kadowaki2022lossy,morita2023random}, which implies that surrogate models are nondeterministic or the uncertainty of trainable parameters are explicitly taken into account. When a given dataset is comprised of data points that belong to a vicinity of a local minimum, we consider that there is a possibility that 
FMA does not work properly. That is, even if we repeat running BBO loops, a series of FM models is trained in a way that the minimum point of every FM model  could be equivalent to such a local minimum. In such a circumstance, we consider that an annealer keeps on selecting the best candidate solutions to be this local minimum point. This indicates that there is a possibility that the candidate solutions searched by FMA get trapped into a local minimum.  We consider that such an aspect is a potential limitation of utilizing FMA.   
\begin{figure*}
\begin{center}
\includegraphics[width=14cm]{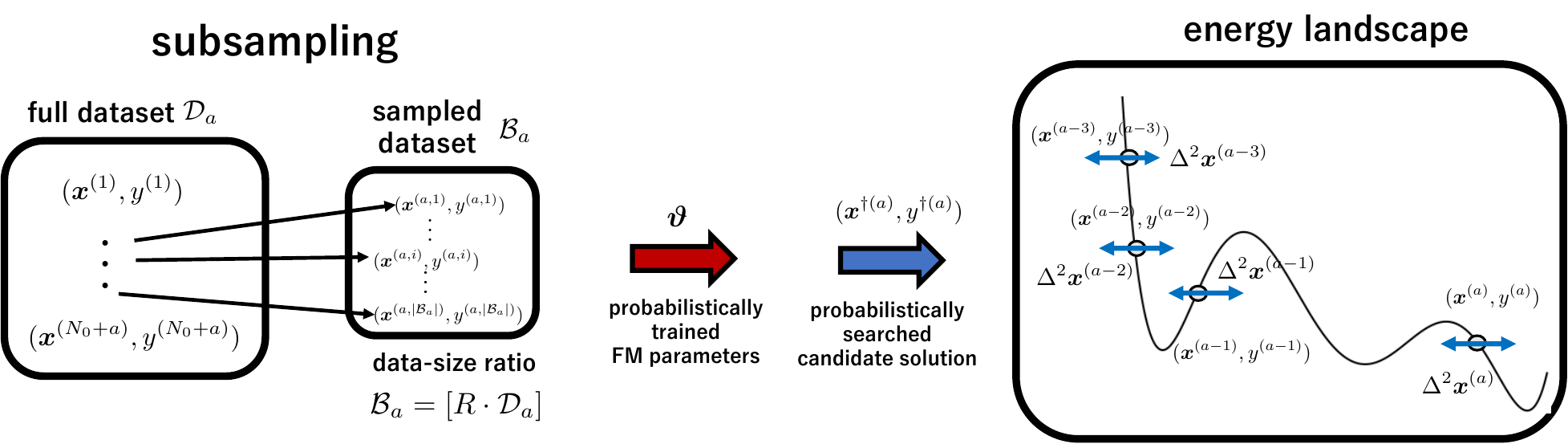}
\end{center}
\caption{(a) Illustration of how to create a sampled dataset $\mathcal{B}_{a}$. It is created by sampling the elements of  $\mathcal{D}_{a}$ according to a probability distribution,
and the relation between the size of $\mathcal{B}_{a}$ ($|\mathcal{B}_{a}|$) and that of $\mathcal{D}_{a}$ ($|\mathcal{D}_{a}|$):  $|\mathcal{B}_{a}| = \left[ R\cdot |\mathcal{D}_{a}|  \right]$ ($0<R< 1$). The data point $(\boldsymbol{x}^{(a,i)}, y^{(a,i)})$ is an $i$th ($i=1,\ldots, |\mathcal{B}_{a}|$) element of $\mathcal{B}_{a}$. Owing to such a probabilistic construction of  $\mathcal{B}_{a}$, the FM function $f_\text{FM}(\boldsymbol{x}; \boldsymbol{\theta}^{(a)})$ is derived probabilistically (FM parameters $\boldsymbol{\theta}$  are trained probabilistically), and  correspondingly, the $a$th best candidate solution $(\boldsymbol{x}^{\dagger(a)}, y^{\dagger(a)}) (= (\boldsymbol{x}^{(a+1)}, y^{(a+1)}))$ is selected from the solution space associated with a deviation.
As a result, we become able to explore a wider range of the energy landscape of $f_\text{BB}(\boldsymbol{x})$ than a case of running FMA as indicated by deviations of the best candidate solutions denoted by symbols $\Delta^2\boldsymbol{x}^{(\alpha)}$ ($\alpha=a-3, \ldots, a$) and the exploration performance gets strengthened.     
}
\label{sampleddataset}
\end{figure*}
 \section{\label{SFMA} Subsampling Factorization Machine Annealing }
 In this section, we explain how to create SFMA and its characteristics. After that, we discuss the advantages of SFMA over FMA and BOCS.
\subsection{\label{sampledsubdataset} Sampled  Subataset} 
   SFMA is constructed by a subdataset that is sampled from a full dataset as shown schematically in Fig. \ref{sampleddataset}, and this is the main difference from FMA. Let us denote a subdataset created in an $a$th BBO loop ($a=1, \ldots, N_\text{ite}$) by  $\mathcal{B}_a=\big{\{} (\boldsymbol{x}^{(a,i)},y^{(a,i)}) \big{\}}_{i=1}^{|\mathcal{B}_{a}|}$, where $|\mathcal{B}_{a}|$ is the size of $\mathcal{B}_{a}$. It  is controlled by a hyperparameter $R$  ($0<R< 1$) via a relation 
$|\mathcal{B}_{a}| = \left[ R\cdot |\mathcal{D}_{a}|  \right]$, where $\left[ \cdots  \right]$ denotes the floor function and $|\mathcal{D}_{a}|$ is the size of the full dataset 
obtained in the  $a$th BBO loop. 
When the FM function $f_\text{FM}(\boldsymbol{x}; \boldsymbol{\theta}^{(a)})$ is trained by an optimization method (e.g., SGD and Adam) with $\mathcal{B}_{a}$, 
the values of $ \boldsymbol{\theta}^{(a)}$ fluctuate or they are trained probabilistically.     
We comment on that such a probabilistic way of training surrogate models is similar to an approach taken in BOCS, which is performed by sampling model parameters according to posterior probability distributions \cite{baptista2018bayesian,koshikawa2021benchmark,matsumori2022application,kadowaki2022lossy,morita2023random}.
Correspondingly, the best solution to  $f_\text{FM}(\boldsymbol{x}; \boldsymbol{\theta}^{(a)})$, which is  $(\boldsymbol{x}^{\dagger(a)}, y^{\dagger(a)}) (= (\boldsymbol{x}^{(a+1)}, y^{(a+1)}))$, also fluctuates. 
Such a fluctuation is going to be a driving force to explore  a broader range of a solution space than an exploration range that can be achieved by FMA  as schematically represented in Fig. \ref{sampleddataset}. 
To support the argument above, we note that it has been investigated in Refs. \cite{keskar2016large,raisa2024subsampling} that the training of machine learning methods using subsampling, which can be called mini-batch methods as well, enables the training of parameters in more nondeterministic ways than those using large-scale subdatasets (large-batch methods). In other words, subsampling (mini-batch) methods are able to explore a broader range of trainable parameter spaces.  
 As a result,  the exploration performance of FMA gets strengthened and
 it is expected that  SFMA exhibits the balanced performance of exploration and exploitation: exploration-exploitation functionality.
  Due to this exploration-exploitation functionality, it is expected that SFMA finds the optimal solution such that in the first half of optimization processes it exhibits  good performance on exploring a wide range of solution space (exploration phase). This is because the dataset size $|\mathcal{B}_{a}|$ is small and the deviation of an FM function as well as its best solution are large.  
 In the second half, it is expected to perform well on searching for the best candidate solutions (exploitation phase) since at this stage the dataset size  at hand is sufficiently big, which indicates that the candidate solutions have been explored sufficiently.
  In Algorithm \ref{pseudocodesamplingFMtraining}, we present a psuedocode for SFMA. The details of standardization given in Algorithm \ref{pseudocodesamplingFMtraining} are described in Appendix \ref{standardization}.     
\begin{algorithm*}[H]
\caption{SFMA }
\label{pseudocodesamplingFMtraining}
\SetKwInOut{Input}{Input}\SetKwInOut{Output}{Output}
\Input{ Initial dataset $\mathcal{D}_0$, BBO iteration number $N_\text{ite}$,  and  ratio $R$.}
\Output{Final dataset $\mathcal{D}_{N_\text{ite}}$ and the optimal solution $(\boldsymbol{x}^{\ast}, y^{\ast}).$}
\BlankLine
\SetKwBlock{Begin}{Procedure}{}
\Begin{
 \tikz[baseline]{\node[fill=black,circle,inner sep=1pt]{}} Generate a dataset $\mathcal{D}_1$ via a single BBO loop with $\mathcal{D}_0$.  \\  
    \tikz[baseline]{\node[fill=black,circle,inner sep=1pt]{}}   
\For{$a \gets 1$ \KwTo $N_\text{ite}-1$}{
1. Create a  subsampled dataset $\mathcal{B}_a$ with $\mathcal{D}_a$  and a ratio $R$. \\  
2. Perform the standardization, $y^{(i)} \to \hat{y}^{(i)}_\text{stand} $, and construct a standardized sampled dataset $\hat{\mathcal{B}}_a$. \\  
3. Train $f_\text{FM}(\boldsymbol{x}; \boldsymbol{\theta}^{(a)})$ with  $\hat{\mathcal{B}}_a$.  \\  
4. Run an annealing process and find the best solution $(\boldsymbol{x}^{\dagger(a)}, y^{\dagger(a)})$.} 
}
\Return{
The final dataset $\mathcal{D}_{N_\text{ite}}$ and the optimal solution $(\boldsymbol{x}^{\ast}, y^{\ast}).$}
\end{algorithm*}

\subsection{\label{compcostandscalability} Computational  Cost and Scalability}
Let us highlight the advantages of SFMA  by making comparisons with FMA and BOCS.  
First and foremost, the significant advantage of SFMA is that it is implementable with lower computational cost than the other two algorithms. That is, when we utilize FMA or BOCS, 
the size of a dataset at hand gets larger as BBO iterations process, which indicates that  the computational cost for machine learning becomes more expensive in the later training processes. In particular, the implementation of BOCS is highly demanding since the computational cost to calculate a posterior distribution per sample is, for instance,
 $\mathcal{O}(p^3)$ or $\mathcal{O}(p\cdot |\mathcal{D}_a|^2)$ ($p = 1+N_\text{bit}+ N_\text{bit}(N_\text{bit}+1)/2$), although it possess  superior exploration performance   \cite{baptista2018bayesian,koshikawa2021benchmark,matsumori2022application,kadowaki2022lossy,morita2023random}. Such an issue becomes prominent for a large-scale problem.
   On the other hand, SFMA has a potential to overcome this issue, which can be done in the following way. 
   When  dataset size as well as $N_\text{bit}$ become substantially large, we take the value of $R$ to be correspondingly small, which implies the $R$ times reduction of the computational cost for training a  surrogate model (FM model) than that  using a full dataset. Furthermore, taking substantially small $R$ is beneficial not only for saving the computational cost for training  an FM model but also for solving a large-scale optimization problem with high accuracy. 
   This is because the deviation of training an FM model enhances by taking $R$ to be  significantly small, and correspondingly, the best candidate solution obtained by an annealer fluctuates substantially.  
   This implies the further amplification of  the exploration performance,  and correspondingly, the exploitation performance is enhanced as well. In this regard, we consider that SFMA has scalability in solving  large-scale problems.
   Note that the exploration performance of SFMA might be weaker than that of BOCS even for a substantially small $R$; however,  we are still able to explore a broad range of a solution space and  SFMA is practically more feasible than BOCS according to the lower computational cost.  

\section{\label{numexps}Numerical Experiments}
Let us demonstrate our numerical experiments on  SFMA and examine whether it  truly possesses the exploration-exploitation functionality  or not
and yields the convergence to the optimal solutions to the given problems with high accuracy. We perform these tasks by benchmarking SFMA against  FMA. 
The problem we choose for our numerical experiments is called lossy compression of data matrices \cite{kadowaki2022lossy,ambai2014spade,yoon2022lossy}, which is an issue or a technique used in, for instance,   
 image recognition, audio data processing, and edge computing. Moreover, lossy compression is related to other techniques such as 
Non-Negative Matrix Factorization and Non-Negative/Binary Matrix Factorization \cite{lee1999learning,long2009rapid,asaoka2020image}, 
 which is utilized in various real-world problems, for instance, image processing and MI \cite{kadowaki2022lossy,lee1999learning,long2009rapid,ambai2014spade,asaoka2020image}.
By showing  the effectiveness of SFMA on such a ubiquitous problem,  {we consider that we can use it as an indicator to  emphasize the utility of SFMA in numerous combinatorial optimization problems in the real world.
In this section, first we explain the problem setting of lossy compression of data matrices. Next, we describe the numerical setup of our experiments.  Finally,  we discuss our numerical results.    
\subsection{\label{probset}Problem Setting }
The objective of this combinatorial optimization problem is to approximately decompose a target $N\times D$ matrix $W$ in terms of  $M$ ($N\times K$ integer matrix) and $C$ ($K\times D$ real matrix) , i.e., $W \simeq MC.$ 
The matrix elements of $M$ take either 1 or -1. By using a pseudoinverse of $M$, which is $(M^\text{T}M)^{-1}M^\text{T},$ the matrix $C$ is approximately re-expressed as  $C \simeq (M^\text{T}M)^{-1}M^\text{T} W,$
and we obtain   $ W  \simeq V(M, W)$ with  $V(M, W) =  M (M^\text{T}M)^{-1}M^\text{T} W$.
To put it another way,  our goal is to construct $V(M, W) $ as a function of $M$ that is sufficiently close to the target matrix $W$. To do this, we introduce two functions of $M$ defined by 
\begin{align}
f_W(M) &=  W  -  V(M, W),  \notag \\
f_{\text{BB},W}(M) &=   ||  f_W(M)      ||_F.
\label{Wdecompfunctions} 
\end{align}
The notation  $ ||  \cdot      ||_F$ in the above equation represents the Frobenius norm defined by  $ || A ||_F = \sum_{i=1}^{n_r} \sum_{i=1}^{n_c}  |a_{i,j}|^2 $, 
where $a_{i,j}$ is an $(i,j)$th matrix element of $A$ and   $n_r$ ($n_c$) is the number of rows (columns) of $A$. 
 Our aim is to find a matrix $M$ that makes  the value of $f_{\text{BB},W}(M)$ as small as possible. Namely, the problem can be reformulated as the combinatorial optimization problem such that the  black-box objective function is $f_{\text{BB},W}(M)$ and the optimal solution  is  $ M^\ast =  \underset{_{M \in \{ -1,1\}^{N \times K}}} {\operatorname{argmin}}  f_{\text{BB},W}(M)$. 
This optimization problem can be reformulated to a binary optimization in the following way. Let
us rewrite $M$ into a form $M=[M_1^\text{T}, \ldots, M_i^\text{T}, \ldots, M_N^\text{T}], $ where $M_i = [m_{i,1},  \ldots, m_{i,j},  \ldots, m_{i,K}]^\text{T}$ with  $m_{i,j}$ denoting 
the $(i,j)$th element of $M$. Since  $m_{i,j} = \pm 1$, it  can be regarded as a spin variable. 
Then, by introducing a new variable $x^M_{i,j} = (1+m_{i,j})/2 $, we obtain an $N_\text{bit} $-dimensional binary vector $\boldsymbol{x}^M = [x^M_{1,1},  \ldots, x^M_{i,j}, \ldots, x^M_{N,K}]^\text{T} \in \{ 0,1 \}^{N_\text{bit} }$
with $N_\text{bit}=N\cdot K$. Therefore, this optimization problem is recast into the black-box optimization problem such that the black-box function is $f_{\text{BB},W}(M)$ in Eq. \eqref{Wdecompfunctions} and  the solution space is the $ 2^{N_\text{bit}} $-dimensional binary solution space.

\subsection{\label{numsetup} Numerical Setup} 
Let us briefly explain the setup of our numerical experiments. For the details, see Appendix \ref{impSFMA}.  
 In our numerical experiments, we create ten types of $W$ matrices (problem instances) that we denote by $W_n$ ($n=0, 1, \ldots, 9$) and set $N_\text{bit}=12,16, 20$.  
 The parameters $(N, D, K)$ are taken to be  $(6, 50, 2), (8, 50, 2)$, and $(10, 50, 2)$ for $N_\text{bit}=12,16,$ and $20$, respectively.
 These $W$ matrices represent model parameters of the shrinked versions of the final fully connected layer of VGG16 convolutional neural network  \cite{simonyan2014very}.
 The repetition number  $N_\text{ite} $ is set to be $N_\text{ite} =C_\text{FM} \cdot N_\text{bit}^2+1$  with  $C_\text{FM}=2,4,6$.
Meanwhile, we take the hyperparameter $k$ [see Eq. \eqref{FMformula1}] to be $k = N_\text{bit}/2 -1.$ In such settings,  
the computational cost is smaller than that for training  general QUBO functions. The sampled datasets $\mathcal{B}_{a}$ are constructed according to uniform sampling. Namely, the elements of $\mathcal{B}_{a}$ are sampled from $\mathcal{D}_{a}$ according to uniform distributions with the probabilities  $1/ |\mathcal{D}_{a}|$ and a fixed random seed. 
 For doing so, we allow duplicate selections.   

Finally, we  discuss the quantities that are calculated in our experiments.    
We calculate $N_\text{samp}$ samples of the datasets $\mathcal{D}_{a} $ that we denote by
$ \big{\{} \mathcal{D}_{a,\alpha}  \big{\}}_{\alpha=1}^{N_\text{samp}}, $ where $\mathcal{D}_{a,\alpha}$ is the dataset  obtained in the $\alpha$th round of the sampling. 
The reason why we perform such  sampling is that the loss functions, which are chosen to be MSE, include fourth-degree terms in the FM parameters $v_{i,l}$ that are responsible for nonconvexities.  
Therefore, when the initial values of $\boldsymbol{\theta}$ differ, in general, the final (updated) values of $\boldsymbol{\theta}$ and the final dataset $\mathcal{D}_{N_\text{ite}}$ differ as well. 
Thus, the benchmarking of these two algorithms must be done by some statistical quantities. In our experiments, first we introduce a quantity defined by $y_{\text{min},\alpha}^{(a)} = \text{min}_{l\in \{1,\ldots, N_\text{in}+a\} } \ y_\alpha^{(l)}$, where $y_\alpha^{(l)} = f_\text{BB}(\boldsymbol{x}_\alpha^{(l)})$ with $\boldsymbol{x}_\alpha^{(l)}$ denoting an $l$th component of the input variable of   $\mathcal{D}_{a,\alpha}$.  
$y_{\text{min},\alpha}^{(a)} $ describes the  minimum value obtained up to the $a$th round of the BBO loop  in the  $\alpha$th sampling. 
Next, we introduce two statistical quantities defined by
$ \bar{y}^{(a)}_\text{min} = (1/N_\text{samp}) \sum_{\alpha=1}^{N_\text{samp}}  y_{ \text{min},\alpha}^{(a)}$ and $R_\text{min,success}^{(a)} = N(y^{(a)}_{\text{min},\alpha}=y^\ast)/N_\text{samp}.$    
The quantity $ \bar{y}^{(a)}_\text{min}  $  describes how 
the candidate solutions obtained by the BBO loops converge to the global minimum (minima) of the given problems as the iteration number $a$ (or the dataset size $|\mathcal{D}_{a}|$) increases.  Meanwhile,
$N(y^{(a)}_{\text{min},\alpha}=y^\ast)$ is the number describing how many  $y_{\text{min},\alpha}^{(a)}$ are equal to the optimal value $y^\ast$. 
In other words, it indicates how many samples out of $N_\text{samp}$ trials  succeeded in finding the optimal solution after BBO loops were performed $N_{\text{it},a} (=|D_a| - |D_0|)$ times.  
Thus, $R_\text{success}^{ (a)}$ is the percentage of $y^{(a)}_{\text{min},\alpha}$  such that  $y^{(a)}_{\text{min},\alpha}=y^\ast$, i.e., the success rate of SFMA (FMA).
 By utilizing $R_\text{min,success}^{(a)} $, we introduce metrics for the exploration performance and the exploitation performance of the algorithms denoted by $N_\text{conv}$ and  $R_\text{success}^\text{final} = R_\text{success}^{(a=N_\text{ite})}$, 
 respectively. The former represents the minimum number of $N_{\text{it},a}$ such that $R_\text{min,success}^{(a)} \geq 0.5$,   whereas the latter describes the final value of $R_\text{min,success}^{(a)}$.
In our numerical experiments,  all the above statistical quantities are computed by taking $N_\text{samp}=30.$  In addition, all these quantities are calculated by rounding them to 17 decimal places in order to avoid floating-point errors. 
\begin{figure*}
\includegraphics[width=16cm]{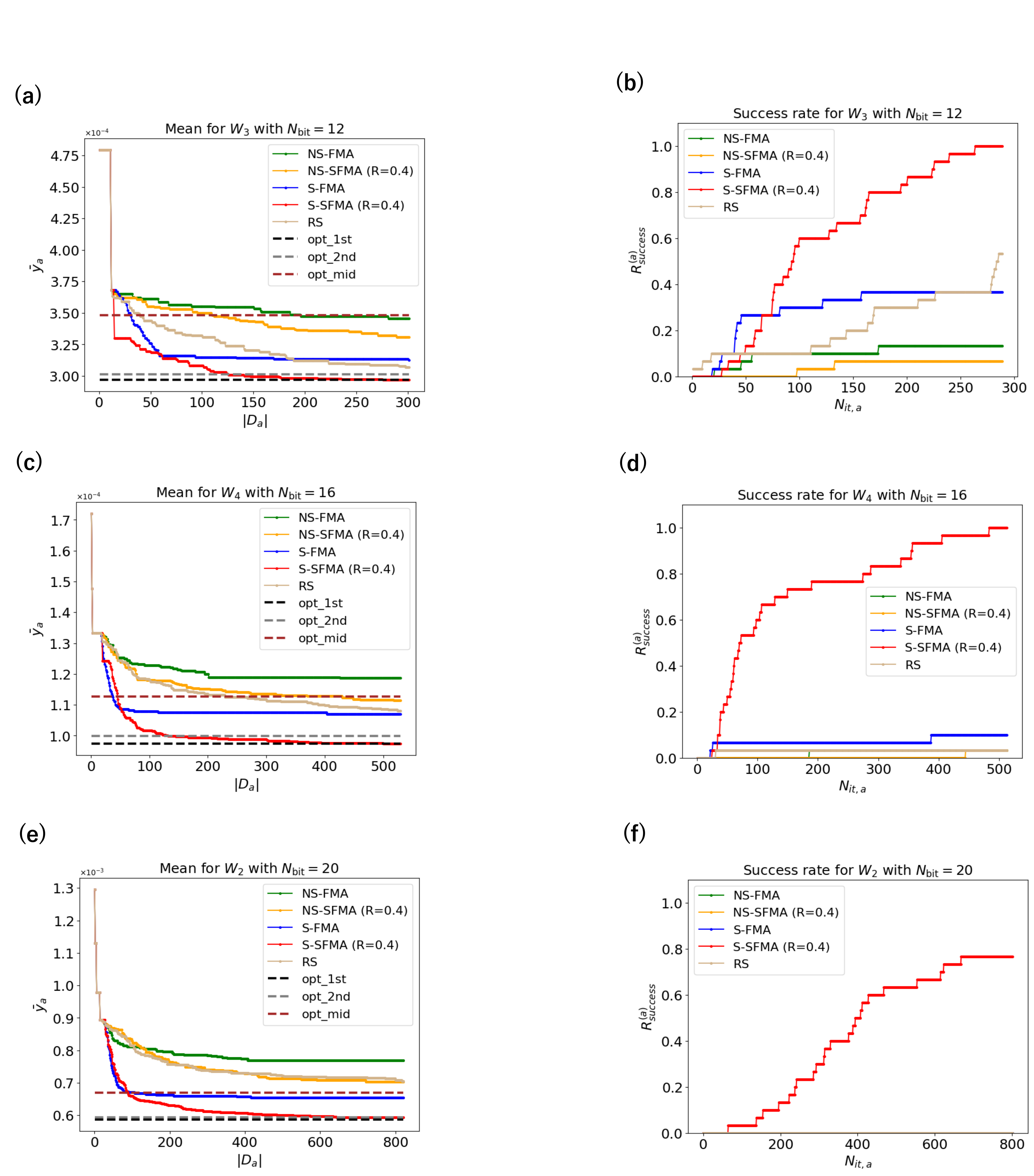}
\caption{Results of  $\bar{y}^{(a)}_\text{min}$ [panels (a), (c), and (e)] and $R_\text{min,success}^{(a)} $ [panels (b), (d), and (f)]. Plots in panels (a) and (b),  (c) and (d), and (e) and (f) display the results for 
$W_3$ with $N_\text{bit}=12$, $W_4$ with $N_\text{bit}=16$, and $W_2$ with $N_\text{bit}=20$, respectively. The iteration number  $N_\text{ite}$ is taken to be $N_\text{ite}=2N_\text{bit}^2+1$. The red (orange) and blue (green) curves represent the results of SFMA and FMA with (without) standardization, respectively. The results of RS are plotted by the tan curves.
  }
\label{mainresultsSAmeansuccessR}
\end{figure*}
\begin{figure*}
\includegraphics[width=15cm]{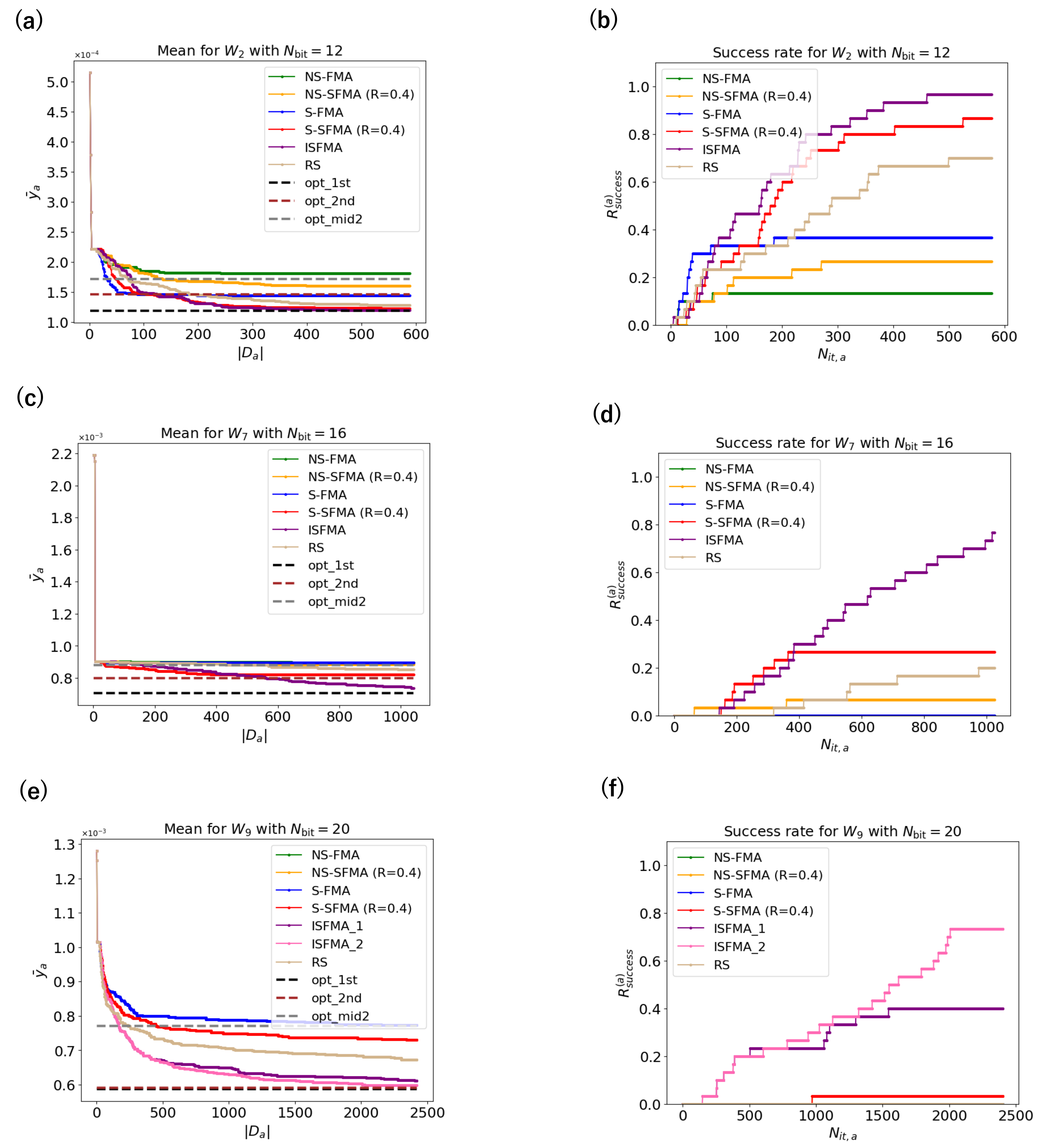}
\caption{Results of  $\bar{y}^{(a)}_\text{min}$ [panels (a), (c), and (e)] and $R_\text{min,success}^{(a)} $ [panels(b), (d), and (f)] of the improved SFMA.
Plots in panels (a) and (b),  (c) and (d), and (e) and (f) are the results for $W_2$ with $N_\text{bit}=12$ and $N_\text{ite}=4N_\text{bit}^2+1$,  $W_7$ with  $N_\text{bit}=16$ and $N_\text{ite}=4N_\text{bit}^2+1$, and $W_9$ with $N_\text{bit}=20$ and $N_\text{ite}=6N_\text{bit}^2+1$, respectively.    For $N_\text{bit}=12$ and 16, the results of the improved SFMA  (ISFMA) are presented by purple curves. On the other hand, for $N_\text{bit}=20$, the two improved algorithms ISFMA$_1$ and ISFMA$_2$ are displayed by purple and pink  curves, respectively.
To make the comparisons between the performance of the improved SFMA and those of the other algorithms, we also plot the results of the nonstandardized FMA (green), the standardized FMA (blue), the nonstandardized SFMA (orange), the standardized SFMA (red), and RS (tan).     
  }
\label{resultsimprovedSA}
\end{figure*}
\begin{figure*}
\includegraphics[width=15cm]{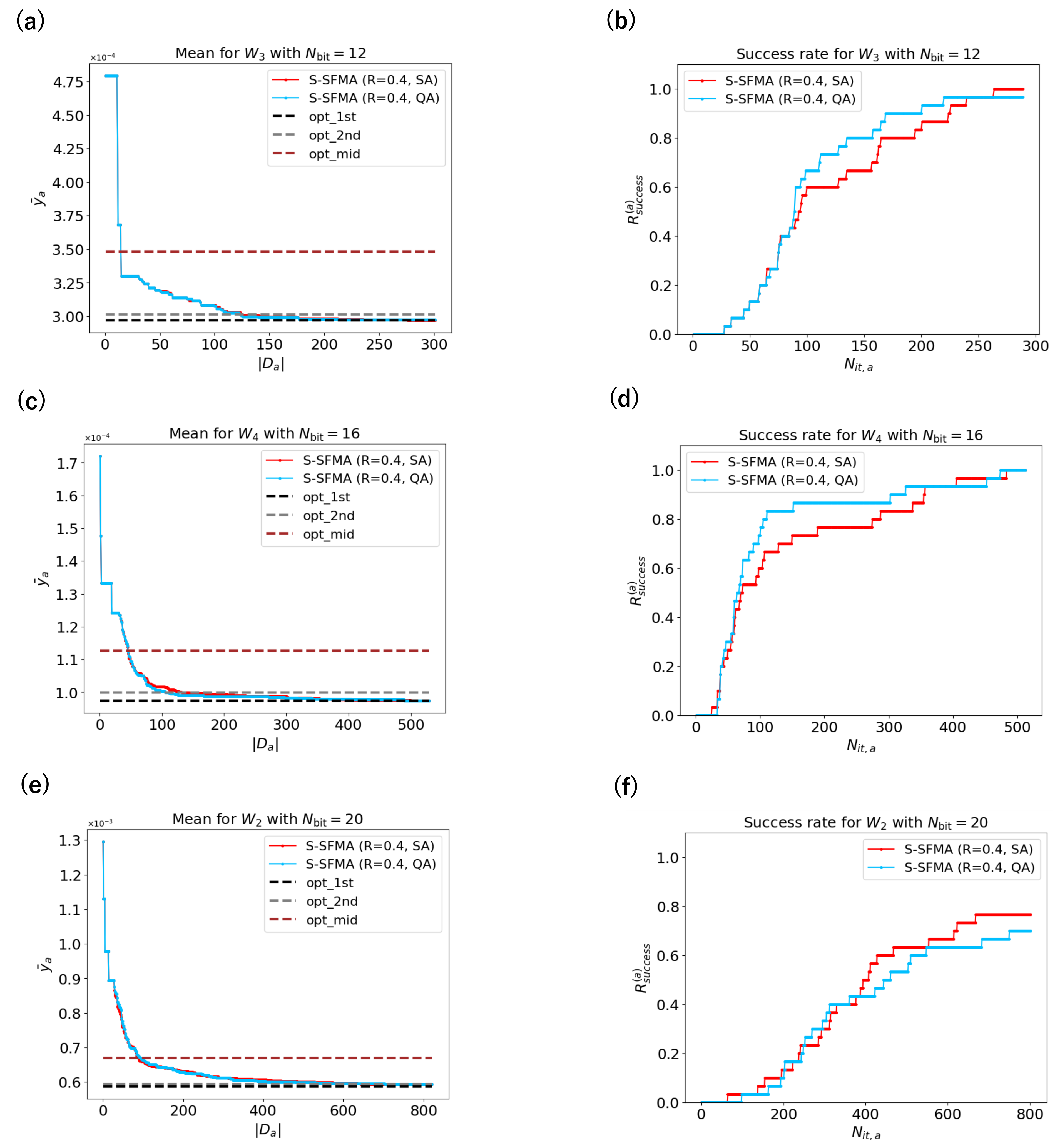}
\caption{Results of  $\bar{y}^{(a)}_\text{min}$ [panels (a), (c), and (e)] and $R_\text{min,success}^{(a)} $ [panels (b), (d), and (f)] of the standardized SFMA  with $N_\text{ite}=2N_\text{bit}^2+1$.
Plots in panels (a) and (b),  (c) and (d), and (e) and (f) are the results for $W_3$ with $N_\text{bit}=12$, $W_4$ with $N_\text{bit}=16$, and $W_2$ with $N_\text{bit}=20$, respectively.  
 The red and light blue curves represent the results for the annealers selected to be SA and QA, respectively. 
  }
\label{resultsSAandQA}
\end{figure*}
  \begin{table*}[t]
   \centering
        \begin{tabular}{c|c|c|c|c|c|c|c|c|c|c|c}
            \diagbox{Algorithm}{Instance}  & $W_0$ & $W_1$  & $W_2$  & $W_3$ & $W_4$  & $W_5$ & $W_6$  & $W_7$ & $W_8$  & $W_9$ & Frequency \\  \hline          
            S-SFMA  & \textbf{396}  & None  & \textbf{393}  &  \textbf{732} & None & None  & \textbf{274} & \textbf{380} & \textbf{321} & None & \textbf{6} \\
             S-FMA  & None  & None & None  & None & None & None  & None & None & None & None & 0 \\
             NS-SFMA  & None  & None  & None  & None & None & None  & None & None & None & None & 0 \\
            NS-FMA  & None  & None  & None  & None & None & None  & None & None & None & None & 0 \\
            RS & None  & None  & None  & None & None & None  & None & None & None & None & 0 \\
        \end{tabular}
           \caption{Data on $N_\text{conv}$ for $N_\text{bit}=20$ and $N_\text{ite} = 2N_\text{bit}^2+1.$ All the annealing processes have been performed by SA.}
        \label{Nconv20}
\vspace{1em}
         \begin{tabular}{c|c|c|c|c|c|c|c|c|c|c|c}
            \diagbox{Algorithm}{Instance}  & $W_0$ & $W_1$  & $W_2$  & $W_3$ & $W_4$  & $W_5$ & $W_6$  & $W_7$ & $W_8$  & $W_9$ & Frequency \\  \hline          
            S-SFMA  & \textbf{24/30}  & \textbf{4/30}  & \textbf{23/30}  & \textbf{15/30} & \textbf{7/30} & \textbf{4/30}  & \textbf{25/30} & \textbf{17/30} & \textbf{24/30} & \textbf{1/30}  & \textbf{10} \\
             S-FMA  &  5/30  & 0/30  & 0/30  & 0/30 & 0/30 & 2/30  & 1/30 & 3/30 & 0/30 & 0/30 & 0 \\
             NS-SFMA  &  0/30  & 0/30  & 0/30  & 0/30 & 0/30 & 0/30  & 0/30 & 0/30 & 0/30 & 0/30 & 0  \\
            NS-FMA  &  0/30  & 0/30  & 0/30  & 0/30 & 0/30 & 0/30  & 0/30 & 0/30 & 0/30 & 0/30 & 0 \\
            RS & 0/30 & 0/30  & 0/30  & 1/30 & 1/30 & 0/30  & 0/30 & 0/30 & 0/30 & 0/30 & 0 \\
        \end{tabular}
          \caption{Data on $R_\text{success}^\text{final}$ for $N_\text{bit}=20$ and $N_\text{ite} = 2N_\text{bit}^2+1.$ All the annealing processes have been performed by SA.}
        \label{finalrate20} 
\end{table*}
\subsection{\label{numresults}Results} 
In this section,  we discuss our numerical experimental results given in Figs. \ref{mainresultsSAmeansuccessR}-\ref{resultsSAandQA} and Tables \ref{Nconv20} and \ref{finalrate20}.
In these figures and tables, we denote the terms nonstandardized and standardized by using acronyms NS and S, respectively; for instance, we call  nonstandardized FMA NS-FMA, while we call  standardized SFMA S-SFMA. In addition to these four algorithms, we present the results of random search (RS) and use them as  indicators to show how well SFMA as well as FMA converge to the optimal solution.  For each $N_\text{bit}$, RS is performed by augmenting  the initial dataset $|\mathcal{D}_0|$ with  $N_{\text{ite}} $  data points, which are selected randomly according to $N_\text{samp}$ different random seeds.  In Figs. \ref{mainresultsSAmeansuccessR}-\ref{resultsSAandQA}, we plot the results for $\bar{y}^{(a)}_\text{min}$ and $R_\text{min,success}^{(a)} $ as functions of the dataset size $|\mathcal{D}_a|$ and  the iteration number of BBO loops $N_{\text{it},a}$, respectively: The plots of $\bar{y}^{(a)}_\text{min}$ ($R_\text{min,success}^{(a)}$) are labeled by a, c, and e (b, d, and f). 
For the plots of $\bar{y}^{(a)}_\text{min}$, we also present three horizontal dashed lines with three colors, black, brown, and gray, in order to present how well SFMA and FMA are approaching the optimal solutions. 
The black, brown, and gray lines represent the output value of the first optimal solution $y_\text{opt,1st}$,  that of the second optimal solution $y_\text{opt,2nd}$, and the approximate optimal values of $f_{\text{BB},W}(M)$ in Eq. \eqref{Wdecompfunctions} ($y_\text{mid}$), which is obtained by an algorithm shown in Ref. \cite{kadowaki2022lossy} (see Eqs. (1)-(5) in  Ref. \cite{kadowaki2022lossy}),  respectively. 
The two values $y_\text{opt,1st}$ and $y_\text{opt,2nd}$ are calculated by exhaustive search (brute force) methods. On the other hand, in Tables  \ref{Nconv20} and \ref{finalrate20}, we present the tabular data on $N_\text{conv}$ and $R_\text{success}^\text{final}$. 
 The last columns denoted by the term frequency indicate how many times each algorithm demonstrated the best performance in convergence (the smallest value of $N_\text{conv}$) and accuracy (the largest value of $R_\text{success}^\text{final}$) across the ten instances. To highlight which algorithm exhibits the best performance in convergence and accuracy,  for each $W_n$ we write the smallest value of $N_\text{conv}$,
  the largest value of $R_\text{success}^\text{final}$, and the largest frequency by the bold texts. When $n_\text{alg}$ ($=1,\ldots, 4$) algorithms and/or RS become tied  ($n_\text{alg}$ algorithms and/or RS exhibit the same smallest values of  $N_\text{conv}$ or the highest values of $R_\text{success}^\text{final}$), we assign a value $1/n_\text{alg}$ [or $1/(n_\text{alg}+1)$] to the frequency of each algorithm except for the cases when the values of $N_\text{conv}$ do not exist as indicated by the term none and when  $R_\text{success}^\text{final}=0$. 
  In addition to  Figs. \ref{mainresultsSAmeansuccessR}-\ref{resultsSAandQA} and Tables \ref{Nconv20} and \ref{finalrate20},   in Appendix \ref{appendixstd}  we demonstrate the analyses of variances and standard deviations of $y^{(a)}_\text{min}$.
In the following, we discuss the results by presenting a section for each figure.
\subsubsection{Results for $N_\text{ite}=2N_\text{bit}^2+1$ via SA} 
In Fig. \ref{mainresultsSAmeansuccessR}, we show the numerical results for $W_3$ with $N_\text{bit}=12$ [panels (a) and (b)], $W_4$ with $N_\text{bit}=16$ [panels (c) and (d)], and $W_2$ with $N_\text{bit}=20$  [panels (e) and (f)]. 
 All these results are obtained by choosing an annealer to be SA with $N_\text{ite}=2N_\text{bit}^2+1$. We explain the setup for the execution of SA in Appendix \ref{setSA}. We present these results since they clearly exhibit the theoretically expected behaviors across the ten $W$ matrices.  The results for the rest of the $W$ matrices are shown in Appendix \ref{otherresults}. Let us explain from the result for the means $\bar{y}^{(a)} $ [Figs. \ref{mainresultsSAmeansuccessR}(a), \ref{mainresultsSAmeansuccessR}(c), and \ref{mainresultsSAmeansuccessR}(e)]. 
We observe in the figure that the means calculated by the standardization approach the  first optimal solution faster than those calculated without the standardization, which implies the effectiveness of the standardization. Such a feature implies the verification of our consideration on the properness of the standardization. 
For the rest of the discussion, let us focus on the results of standardized SFMA and FMA and analyze their behaviors.
 Basically, we observe the same characteristics across all $N_\text{bit}$. That is, in the first half of the BBO processes  the red curves (S-SFMA) take the equivalent or larger values than those of the blue curves (S-FMA) in certain ranges. 
 On the other hand, in the second half S-SFMA exhibits the lower values at all  $|\mathcal{D}_a|$ and approaches the first optimal solution with significantly higher accuracy. 
 Correspondingly, in Figs. \ref{mainresultsSAmeansuccessR}(b), \ref{mainresultsSAmeansuccessR}(d), and \ref{mainresultsSAmeansuccessR}(f) we observe that in the first half the values of $R_\text{min,success}^{(a)}$ calculated by S-SFMA take the equivalent or smaller values  in certain ranges whereas in the second half 
 they are larger at  all $N_{\text{it},a}$.  We interpret such contrastive characteristics of the SFMA and FMA in the following way.
  In the first half, SFMA exhibits the larger deviation in searching the optimal solution since it is in the exploration phase with higher performance than the exploration performance of FMA.  In the second half, the candidate solutions are explored sufficiently by SFMA and it exhibits higher explotation performance. As a result, SFMA finds the optimal solution more efficiently and effectively than FMA. 
  This consideration is supported by Tables \ref{Nconv20} and \ref{finalrate20}, which present the data on $N_\text{conv}$  and $R_\text{success}^\text{final}$  for $N_\text{bit}=20$, respectively. 
  From these tables, we clearly observe that SFMA shows the advantages over the other algorithms in both speed (convergence) and accuracy.     

\subsubsection{Results for $N_\text{ite}=4N_\text{bit}^2+1, 6N_\text{bit}^2+1$ via SA}
Next, we discuss the results in Fig. \ref{resultsimprovedSA}. As indicated by Fig. \ref{mainresultsSAmeansuccessR}, Figs. \ref{otherresultsWsspin12mean}-\ref{otherresultsWsspin20successrate} in  Appendix \ref{otherresults}, and  Tables \ref{Nconv20} and \ref{finalrate20}, the ways the algorithms converge to the optimal solutions do not show the similar behaviors with respect to $W_n$. 
Specifically,  S-SFMA converge to the optimal solutions with high accuracy (large $R_\text{success}^\text{final}$) for certain cases including the ones shown in Fig. \ref{mainresultsSAmeansuccessR}, whereas it does not (small $R_\text{success}^\text{final}$) for some cases. Furthermore, the values of $N_\text{conv}$  of S-SFMA does not exist for some instances.  The convergence performance as well as the accuracy of SFMA can be improved by, for instance, increasing the value of $N_\text{ite}$ and/or taking the value of $R$ to be smaller than 0.4. By controlling the values of $N_\text{ite}$ and $R$ coherently,  SFMA can be improved in various ways and we investigate them  in the following.  
In  Fig. \ref{resultsimprovedSA}, we present the results for $W_2, W_7$, and $W_9$ with $N_\text{bit}=12,16,$ and $20$, respectively: the results  for other $W$ matrices are discussed in Appendix \ref{improvedSFMA}. For  $N_\text{bit}=12$ and 16, the parameters $N_\text{ite}$ and $R$ are taken to be $N_\text{ite}=4N_\text{bit}^2+1, R=0.1$.  
We call these algorithms ISFMA.
For $N_\text{bit}=20,$  we make two types of improvements. The first one is done by taking $N_\text{ite}=6N_\text{bit}^2+1, R=0.1$ 
and the second one is done by sequentially running two different types of SFMA.  In the first part, we perform SFMA with $N_\text{ite} = N_\text{bit}^2$ and $R=0.1$, whereas in the second part we run it by taking $N_\text{ite} = 5N_\text{bit}^2$ and $R=0.01.$  
For convenience, we call the former algorithm ISFMA$_1$ whereas the second one ISFMA$_2$.     
We benchmark them against the other four different algorithms (standardized and nonstandardized FMA and SFMA with $R=0.4$) and RS in convergence and accuracy.
These algorithms are performed by augmenting  the datasets whose sizes are equal to $2N_\text{bit}^2+1$ (the results shown in Figs. \ref{otherresultsWsspin12mean}-\ref{otherresultsWsspin20successrate})  with  additional  datasets  obtained by $2N_\text{bit}^2 $ ($4N_\text{bit}^2 $) times of BBO loops for $N_\text{bit}=12$ and 16 ($N_\text{bit}=20$).
 Let us explain from  the cases of $N_\text{bit}=12$ [panels (a) and (b)] and $16$ [panels (c) and (d)].  For both cases, we observe that  ISFMA  (purple curve) exhibits  the smallest values of $N_\text{conv}$ and the highest values of $R_\text{success}^\text{final}$. 
          
 Finally, we discuss the result for $N_\text{bit}=20$ [panels (e) and (f)].  
  What we observe from the figures is that  ISFMA$_1$ (purple curve) and ISFMA$_2$ (pink curve) approach the optimal solution  faster than any other four algorithms and RS. 
 In the following, let us focus on the results of ISFMA$_1$ and ISFMA$_2$ and analyze their characteristics more in detail. 
 Note that both of these algorithms are performed such that the second parts of BBO calculations are carried out  using  the same dataset obtained  by running SFMA with $N_\text{ite} = N_\text{bit}^2$ and $R=0.1$ (the first part of BBO calculation). The results demonstrate that in the first half  the mean values $\bar{y}^{(a)} $ of ISFMA$_1$ are smaller than those of ISFMA$_2$ 
 in certain ranges.  In the second half, the mean values $\bar{y}^{(a)} $ of ISFMA$_2$ are  always smaller than those of ISFMA$_1$.
 Correspondingly, in the first half, the success rates $R_\text{min,success}^{(a)} $ of ISFMA$_2$ are equal to or smaller in certain ranges,       whereas  
  in the second half the values of $R_\text{min,success}^{(a)} $ of ISFMA$_2$ become equivalent or  larger at  all $N_{\text{it},a}$. 
  The final value of the success rate $R_\text{success}^\text{final}$ of ISFMA$_2$ is equal to 22/30 whereas that of ISFMA$_1$  is $12/30$, which implies that the accuracy of ISFMA$_2$  is about  twice as large as that of  ISFMA$_1$. 
We consider that such a significant difference between these two algorithms has been generated due to how the FM models have been trained  in the second part.  That is,
in ISFMA$_2$  the FM models have been trained by taking $R=0.01$,  which is ten times smaller than the value of $R$  taken in ISFMA$_1$.
 We consider that such a way of training has enabled ISFMA$_2$ to explore a significantly  wider range of the solution space, which leads the enhancement of the exploration performance of ISFMA$_2$
and the exploitation performance as well.  Moreover,  ISFMA$_2$  exhibits the value of $N_\text{conv}$ while ISFMA$_1$ does not.
   As a result,  both the performance of exploration and exploitation of  ISFMA$_2$ are significantly higher than the other algorithms.   
   We regard this result as the reflection of the potential scalability of SFMA discussed in Sec. \ref{compcostandscalability}. 

\subsubsection{Results for $N_\text{ite}=2N_\text{bit}^2+1$ via QA}
 Finally, let us discuss the results in Fig. \ref{resultsSAandQA}. These are the results for $\bar{y}^{(a)} $ and $R_\text{min,success}^{(a)} $ computed by running S-SFMA via
 both SA (red) and QA (light blue). Note that the red curves are the same ones presented in Fig. \ref{mainresultsSAmeansuccessR}.
 The numerical setup for running QA is explained in Appendix \ref{setQA}.
   We observe  in  Fig. \ref{resultsSAandQA}  that the results obtained by the two annealers quantitatively demonstrate almost the same behaviors.  In our experiments, we have not observed the explicit outperformance of QA over SA in speed and/or accuracy (quantum advantage).  
 Meanwhile, for instance, in Refs. \cite{denchev2016computational,albash2018demonstration}, it has been highlighted that for certain optimization problems QA might be more advantageous than SA. In addition,  it has been studied in Ref. \cite{pirnay2024principle} that a quantum computer exhibits  a superpolynomial quantum advantage over a classical computer in finding an approximate solution to combinatorial optimization problems.   
 When $N_\text{bit}$ is sufficiently larger than the values taken in our experiments, the complexity of a Hilbert space, which is a resource of quantum computing, becomes more enhanced,  and we expect that there is a possibility that we observe the supremacy of QA such that  
 it demonstrates advantages  in speed and/or accuracy over SA. We would like to address such an issue in our future work. 

\section{\label{conclusion}Conclusion and Outlook}
In this work, we have developed the algorithm that enables us to  solve black-box optimization problems effectively and efficiently by devising the training processes: SFMA.  
FMA  is one of the methodologies for solving black-box optimization problems; however, it has a potential limitation such that its performance on exploring candidate solutions may be insufficient since 
FM models are trained by a point-estimation  approach.
From this perspective, we have improved FMA  by strengthening  its exploration performance. 
Specifically, we have trained FM models  by sampled subdatasets.
Since FM models are trained in such probabilistic ways, we become able to explore broader ranges of solution spaces, which implies the amplification of the exploration performance of FMA.  
As a result, SFMA exhibits the exploration-exploitation functionality. We have conducted the numerical experiments to verify whether SFMA truly exhibits the exploration-exploitation functionality or not and to benchmark it against FMA in speed (convergence) and accuracy.   Consequently,  we have verified that SFMA certainly exhibits the exploration-exploitation functionality and is superior to FMA in both speed  and accuracy. Furthermore, we have observed the potential scalability of SFMA by setting the value of $R$ to be correspondingly small according to the problem size.
   We expect that the advantages of SFMA, low computational cost and scalability,  are going to be powerful tools for tackling  complex problems in the real world. 
   
   Let us further highlight the superiority of SFMA  with respect to the computational cost. 
  In this work,  we have amplified the exploration performance by subsampling.   
   Meanwhile, we are able to strengthen  the exploration performance in other ways. 
   For instance, we can accomplish this by running FMA and BOCS alternatively. Specifically,  first we run BOCS and later we execute FMA. Moreover,  BOCS can be replaced with another stochastic algorithm. 
   Although we can take such alternative approaches, we consider that using the subsampling method  is more beneficial.
   This is because  the computational cost for running stochastic algorithms like BOCS typically increases significantly as the dataset size and/or the dimension of a solution space become larger.
   On the other hand, SFMA can be performed with  low computational cost even for a large dataset by setting the value of $R$ to be correspondingly small, although its exploration performance might be weaker than other stochastic algorithms.    Another approach to accomplish establishing the algorithms possessing the exploration-exploitation functionalities is, for instance, to enhance the exploitation performance of stochastic algorithms.  We would like to address this problem in our future work by examining various stochastic algorithms.  
   Finally, we would like to point out that the subsampling approach works for training other types of machine learning models. 
   With this taken into account,  the investigation of the machine learning model trained by a sampled subdataset that demonstrates the best performance on a given problem would also be an interesting future direction.   In relation to this, the investigation of other sampling methods would be another important theme to explore. These could be, for instance,  the creation of subdatasets using clustering methods and that comprised of small and large values of  objective functions. 
   Furthermore, the optimization of the hyperparameter $R$ would be another interesting direction. 
 
In this work, we have developed the effective algorithm for black-box optimization problems by devising  dataset modeling, i.e.,  step (ii) in Fig. \ref{BBOschematic}.
By developing powerful  method for each of the other two basic steps and incorporating them into SFMA coherently, 
we expect that we become able to run SFMA more effectively and efficiently.  
Furthermore, the development of powerful algorithms for BBO that are carried out by  gate-based quantum optimization algorithms \cite{au2024quantum,farhi2014quantum,hadfield2019quantum,blekos2024review,abbas2024challenges} would be another interesting and important direction. 

Recently, intensive research and development in both machine learning (artificial intelligence) and quantum technology including both hardware engineering and algorithmic development has been progressing. It is expected to continue toward the development of  algorithms that are capable of addressing complex problems in the real world and  large-scale (both classical and quantum) devices that enable us to leverage such algorithms. 
We expect that our results pave the way for accomplishing these goals and lead to
the creation of next-generation technologies that are going to be the basic building blocks for accelerating industrial developments. 

\section*{Acknowledgments} 
 We thank the members of G-QuAT Quantum Application Team for having a fruitful discussion on our work.
This work was done for Council for Science, Technology and Innovation (CSTI), Cross-ministerial Strategic Innovation Promotion Program (SIP), “Promoting Application of Advanced Quantum Technologies to Social Challenges" (Project management agency: QST) under Grant No.  JPJ012367. 

\section*{Data Availability} 
The data files presented in this paper can be obtained at Zenodo \cite{sfmazenodo}.
  
\widetext 
\appendix
\section{\label{standardization} Standardization Method}
 We explain how to construct the normalized sampled subdatasets $\hat{\mathcal{B}}_a$ in Algorithm \ref{pseudocodesamplingFMtraining} . The subdatasets $\hat{\mathcal{B}}_a$ are formulated by the standardized output variables defined by \cite{suh2022convex,mehta2019high,murphy2022probabilistic,bishop2023deep,kadowaki2022lossy} 
\begin{align}
\hat{y}^{(i)}_\text{stand}  &= \frac{ y^{(i)} -\bar{y}}{ \sqrt{\Delta^2 y} \times N_\text{bit}}.
\label{standardizedBBF}
\end{align}
 The quantities $\bar{y}$ and $\Delta^2 y$ in the above equation are defined by
\begin{align}
\bar{y} &= \frac{1}{N_\mathcal{G} }   \sum_{y \in  \mathcal{G}} y , \notag \\
\Delta^2 y  &= \frac{1}{N_\mathcal{G} }  \sum_{y \in  \mathcal{G}} (y - \bar{y})^2.
\label{standardyaverageandvariancen}  
\end{align} 
where $\mathcal{G}$ is an ensemble of pairs of input and output variables constructed randomly from a full dataset and  $N_\mathcal{G} $ is the number of  such pairs  included in $\mathcal{G} $. 
In our numerical experiments, we take  $N_\mathcal{G} = 5\cdot N_\text{bit}.$
The reason why we use the standardized output variables $\hat{y}^{(i)}_\text{stand} $ in Eq. \eqref{standardizedBBF} instead of the bare output variables $y^{(i)} $ to train FM models is because when the values of $y^{(i)} $ are considerably small, there are possibilities that annealing processes do not work effectively.
Let us discuss more in detail for both SA and QA. In case of SA, when output values $y^{(i)} $ are too small this effectively corresponds to taking high inverse temperatures, which means that the annealing processes tend to show random behaviors. Such randomness destabilizes the convergence to the optimal solutions, which implies that the annealing processes do not work effectively. We consider that such destabilization of  convergence is prominent for large $N_\text{bit} $ since the dimension of a solution space is significantly big. Next,  let us discuss the effectiveness of the standardization on QA. We consider that the standardization  is also effective on QA because of the following reason. When bare output variables are substantially small, then  the energy gaps between the ground state energies and the first excited energies are considerably small as well. In such cases, the performance of QA gets degraded significantly. By performing the standardization, the magnitudes of such energy gaps are effectively enhanced, and correspondingly, the performance of QA is amplified. 
 Therefore, the standardization is also effective for performing QA with high accuracy.    
\section{\label{impSFMA} Implementation of SFMA} 
In this section,  we describe the implementation of SFMA in our numerical experiments. 
First, we explain the numerical procedure of SFMA, which is represented schematically in Fig. \ref{nprocSFMA}. 
 After that, we explain the setups for running SA and QA. 
 All the classical computations in our numerical experiments are performed on a  MacBook Air (Sonoma 14.2) that consists of  Apple M2 (8-core CPU) and 24 GB of memory. 
\begin{figure*}
\begin{center}
\includegraphics[width=14cm]{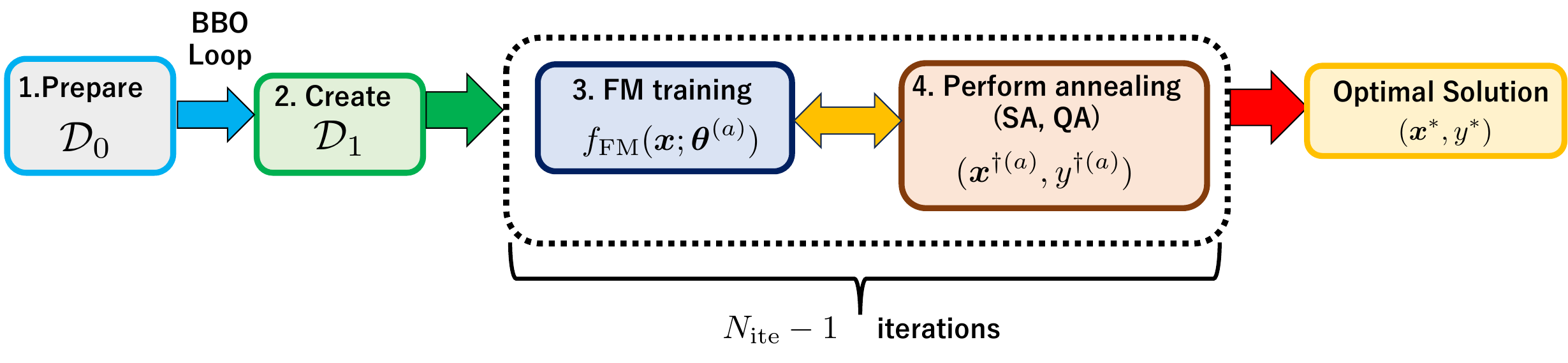}
\end{center}
\caption{Numerical procedure of SFMA comprised of four basic steps.  First, we prepare an initial dataset  $\mathcal{D}_0$.  
Second, we create  $\mathcal{D}_1$. Third, we train an FM model $f_\text{FM}(\boldsymbol{x}; \boldsymbol{\theta}^{(a)})$.
Fourth, we run annealing (SA or QA) to search for the optimal solution to  $f_\text{FM}(\boldsymbol{x}; \boldsymbol{\theta}^{(a)})$ denoted by
$(\boldsymbol{x}^{\dagger(a)}, y^{\dagger(a)})$. Steps three and four are performed  iteratively  $N_\text{ite}-1$ times.
 }
\label{nprocSFMA}
\end{figure*}
\subsection{\label{numprosSA}  Numerical Procedure } 
The numerical procedure of SFMA is comprised of four basic steps. 
First,  we prepare an initial dataset $\mathcal{D}_0$ with $ |\mathcal{D}_0|=N_\text{bit}$ for each problem instance  generated by randomly selecting $N_\text{bit}$ sets of input and output variables using a fixed random seed.
 Second, for each problem instance we create a full dataset $\mathcal{D}_1$ by running a single BBO loop. We perform this in two ways, the training of FM models with and without standardization. The reason we start running SFMA  not with $\mathcal{D}_0$ but with $\mathcal{D}_1$ is because we conduct the benchmarking tests in a way that we compare the performance of SFMA and that of FMA after the initial dataset $\mathcal{D}_0$, which is the common given dataset, is used for both cases. In the rest, SFMA is performed $N_\text{ite}-1$ times. Since each iteration is performed by the same way in terms of the basic three steps in  Fig. \ref{BBOschematic}, 
   we describe the following description as the explanation of the procedure of running the $a$th ($a=2, \ldots, N_\text{ite}$) BBO loop.  
 Third, we train the FM models $f_\text{FM}(\boldsymbol{x}; \boldsymbol{\theta}^{(a)})$ with the datasets $\mathcal{D}_{a-1}$. 
 We do this by optimizing MSE using Adam with all the samples of the datasets (full batch).  For every calculation, the learning rate $\eta$ and the epoch number $N_\text{epoch}$ are taken to be $\eta = 0.01$ and $N_\text{epoch}=200$, respectively. We note that for each iteration, each component  of the FM  parameters $\boldsymbol{\theta}^{(a)}$ is trained by initially generating a random number obeying a normal distribution with its mean and variance equal to 0 and $\Delta^2 \bar{y}$ given in Eq. \eqref{standardyaverageandvariancen}, respectively. 
 We take the variance of the normal distribution to be $\Delta^2 \bar{y}$ since the FM model parameters  $\boldsymbol{\theta}^{(a)}$ have the same orders with the black-box functions. In other words, we also rescale $\boldsymbol{\theta}^{(a)}$ as we do in the standardization: In contrast, in cases of the nonstandardized SFMA and FMA we do not perform rescaling.  
Such a normal distribution is constructed by a fixed random seed.  All these machine learning calculations are carried out using PyTorch version 2.4.1 and Python version 3.12.7.  
 Fourth, we run annealing processes (SA or QA) using D-Wave Ocean SDK version 8.1.0. 
 \subsection{\label{setSA} Setup for SA}
Let us describe the setup for SA. The SA calculations are performed using the library called
neal.SimulatedAnnealingSampler, which is supported by D-Wave Ocean SDK \cite{dwaveoceansdk}. We set two parameters,  num$\_$reads  and num$\_$sweeps,   and a fixed random seed. 
For convenience, hereinafter let us denote num$\_$reads  and num$\_$sweeps by $n_\text{reads}$ and $n_\text{sweeps}$, respectively.
 The parameter  $n_\text{reads}$ represents the repetition number of a definite single annealing procedure and  $n_\text{sweeps}$ describes the number of Markov chain Monte Carlo steps.  In our experiments, we take $(n_\text{reads},n_\text{sweeps})=(10,100)$ and the rest of the parameters are set to their default values.   We note that we have verified that the annealing processes under these values have enabled the acquisitions of the optimal solutions to  $f_\text{FM}(\boldsymbol{x}; \boldsymbol{\theta}^{(0)})$ (FM function obtained by $\mathcal{D}_0$) trained in both cases, with and without the standardization procedure.    Thus, the SA processes work properly for both cases under these values of $n_\text{reads}$ and $n_\text{sweeps}$.
\subsection{\label{setQA} Setup for QA}
Let us explain the setup for running QA.    The quantum annealing calculations have been performed by the quantum device called Advantage$\_$system 6.4,  which is supported by D-Wave Systems, Inc. \cite{dwaverealquantumdevice}. 
  In order to implement the FM models on it, the logical qubits are mapped onto the physical qubits, which is so-called minor embedding. 
  All the minor-embedding processes are carried out by utilizing the function called EmbeddingComposite, which is provided in D-Wave Ocean SDK \cite{dwaveoceansdk}.  
    In addition, as similar to the cases of SA, a definite QA process is run multiple times independently, which is characterized by a parameter num$\_$reads: In the following, we denote it by $n_\text{reads}$. 
    For every experiment, we take $n_\text{reads}=50$ and the rest of the parameters are set to their default values. Furthermore,
    we have numerically verified that QA properly finds the optimal solutions to $f_\text{FM}(\boldsymbol{x}; \boldsymbol{\theta}^{(0)})$ under this parameter setting. 

\begin{figure*}
\begin{center}
\includegraphics[width=16cm]{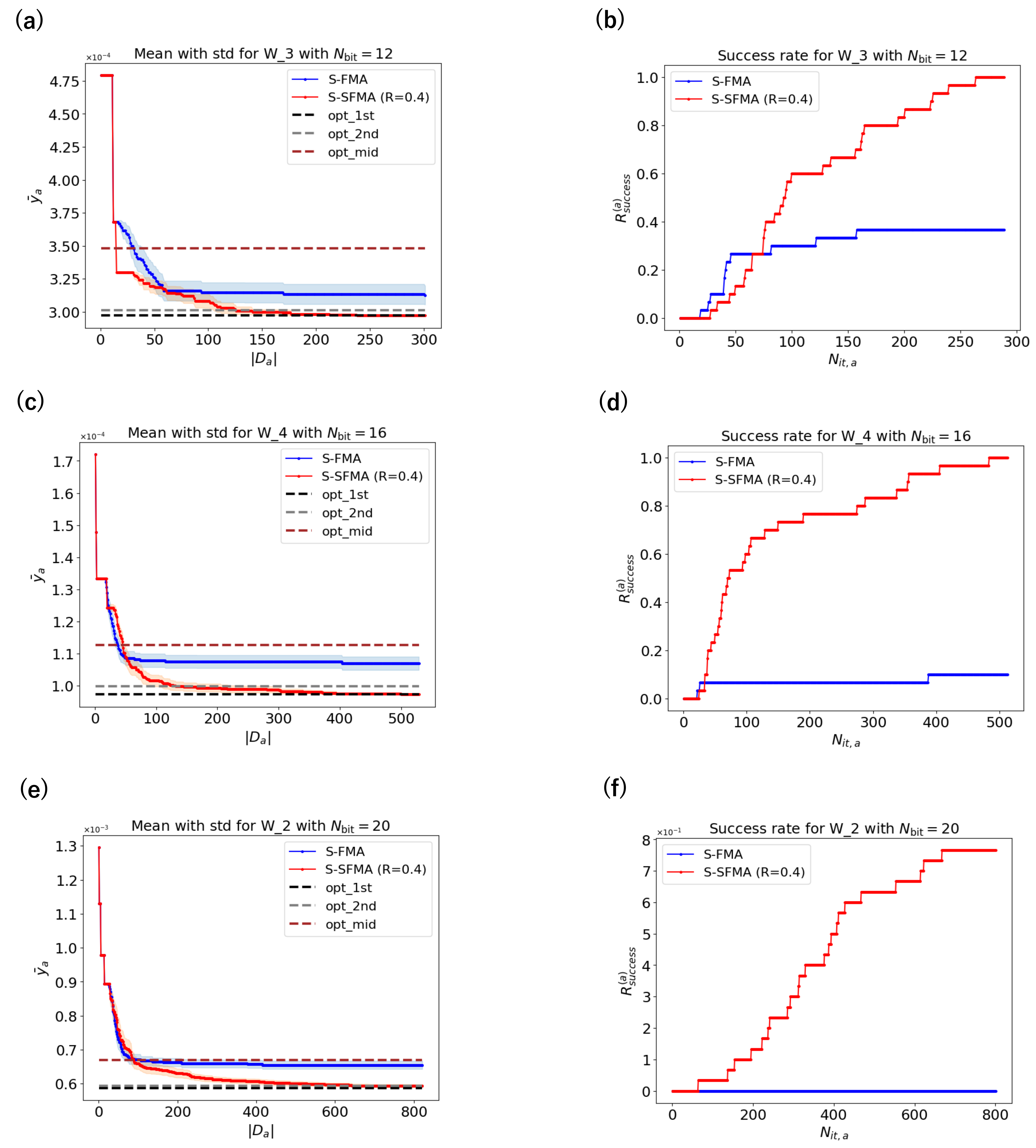}
\end{center}
\caption{Results of $\bar{y}^{(a)} $ associated with standard deviations $ \sqrt {\sigma^2 }$ [panels (a), (c), and (e)] and $R_\text{min,success}^{(a)} $ [panels (b), (d), and (f)] calculated by S-SFMA  (red) and S-FMA (blue).  
The plots in  panel (a) and (b), (c) and (d), and (e) and (f) are the results for  $W_3$ with $N_\text{bit} =12$, $W_4$ with $N_\text{bit} =16$, and $W_2$ with $N_\text{bit} =20$, respectively. 
 $N_\text{ite}$ is taken to be $N_\text{ite}=2N_\text{bit}^2+1$ for every case.
 }
\label{meanstdrate}
\end{figure*}

\section{\label{appendixstd}  Analyses of Standard Deviations}
In Sec. \ref{numresults}, we have discussed the results for  the means $\bar{y}^{(a)} $ and the success rates $R_\text{min,success}^{(a)} $.
In addition, let us analyze our results of variances defined by   
\begin{align}
  \sigma^2 =    \frac{1}{N_\text{samp}} \sum_{\alpha=1}^{N_\text{samp}} \left(
 y_{ \text{min},\alpha}^{(a)} -  \bar{y}^{(a)}_\text{min}  \right)^2. 
\label{minstd} 
\end{align}  
 In the following, we investigate the behaviors of the associated standard deviations $ \sqrt { \sigma^2 }$ given by Eq. \eqref{minstd} by comparing to those of the success rates $R_\text{min,success}^{(a)} $.  
 In Fig. \ref{meanstdrate}, we display the results  of S-SFMA and  S-FMA by red and  blue curves, respectively. The plots in panels (a) and (b), (c) and (d), and (e) and (f) denote the results for $W_3$ with $N_\text{bit} =12$, $W_4$ with $N_\text{bit} =16$, and $W_2$ with $N_\text{bit} =20$, respectively. For every calculation, we take the BBO iteration number  to be $N_\text{ite}=2N_\text{bit}^2+1$. 
In panels (a), (c), and (e), we display the plots of the means $\bar{y}^{(a)}_\text{min}$ with shaded regions representing 95$\%$ confidence intervals  defined by 
$\big{[} \bar{y}^{(a)}-1.96 \sqrt { \sigma^2/N_\text{samp}}, \bar{y}^{(a)}+1.96 \sqrt { \sigma^2/N_\text{samp}} \big{]}$. 
On the other hand, we present  the results for $R_\text{min,success}^{(a)} $ in panels (b), (d), and (f). 
Note that $\bar{y}^{(a)}_\text{min}$ and $R_\text{min,success}^{(a)} $ shown here are the same ones  in Fig. \ref{mainresultsSAmeansuccessR}. 
From the results in panels (a), (c), and(e), we observe that, for all $N_\text{bit}$,  when the dataset sizes $|\mathcal{D}_a|$ increase and get closer to $N_\text{tot}$, the standard deviations become sufficiently small or approach zero for the cases of S-SFMA. 
In contrast,  the standard deviations remain finite for the cases of S-FMA. Correspondingly,   the results in panels (b), (d), and (f) demonstrate that for the cases of S-SFMA,  the values of $R_\text{min,success}^{(a)} $ increase as $N_{\text{it},a}$ get larger 
and become equal to or almost equal to 30/30 at $N_{\text{it},a} =N_\text{ite}$. For the cases of S-FMA, the values of $R_\text{min,success}^{(a)} $ get larger as $N_{\text{it},a}$ increase; however, the final values $R_\text{min,success}^{\text{final}} $ 
are equal to or smaller than 11/30. As a result,  the standard deviations are zero (sufficiently small) when we achieve the optimal solutions in all (almost all) $N_\text{samp}$ ($=30$) trials whereas they take moderate values when  the optimal solutions have been obtained in the portions of the $N_\text{samp}$ trials as expected.  
\begin{figure*}
\begin{center}
\includegraphics[width=17cm]{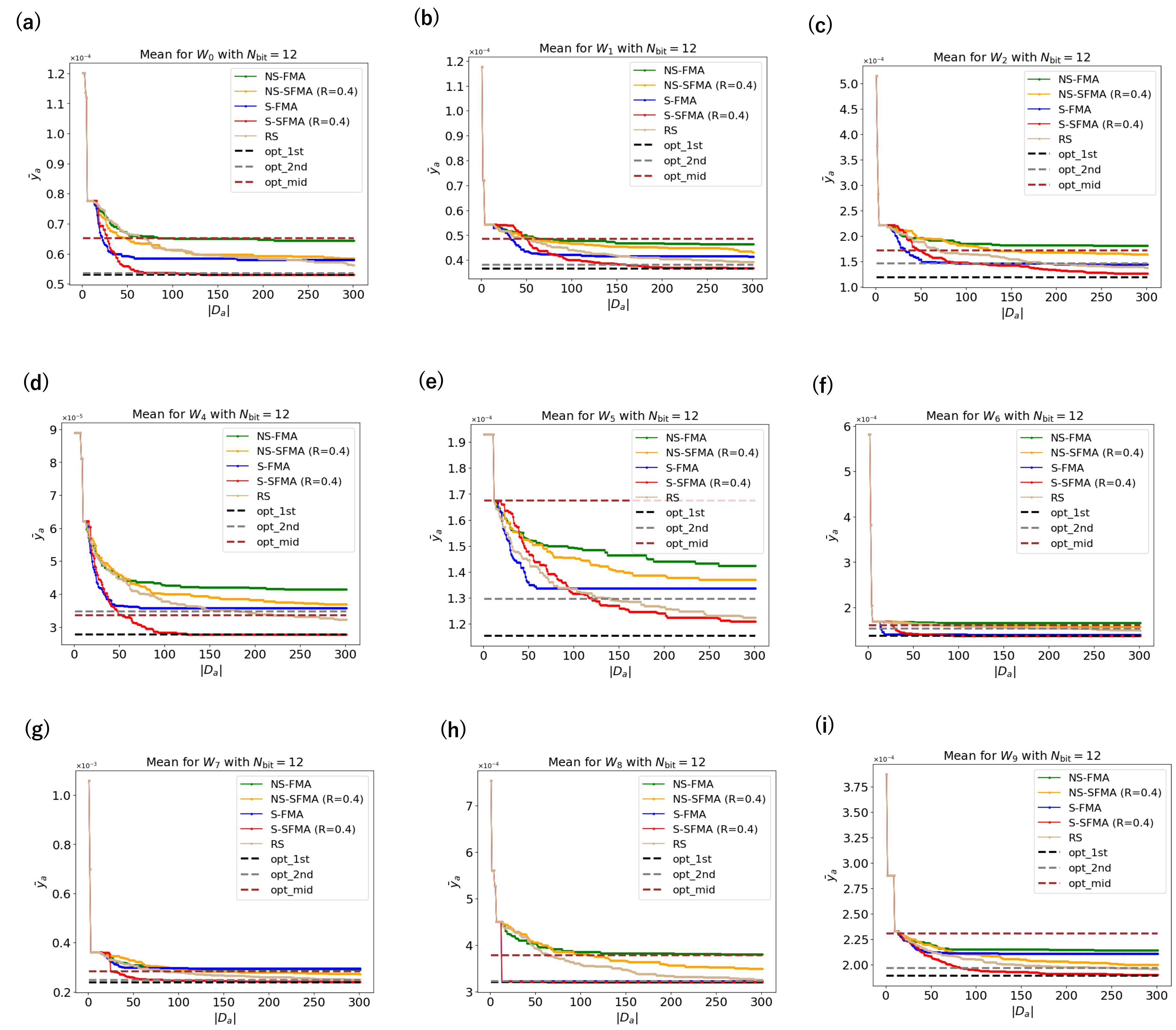}
\end{center}
\caption{Results for  $\bar{y}^{(a)} $ with $N_\text{bit} =12$ and $N_\text{ite}=2N_\text{bit}^2+1$. Plots in panels
 (a)-(i) are the ones for $W_n$ with $n=0,1,2,4,5,6,7,8,$ and $9$, respectively. }
\label{otherresultsWsspin12mean} 
\end{figure*}
\begin{figure*}
\begin{center}
\includegraphics[width=17cm]{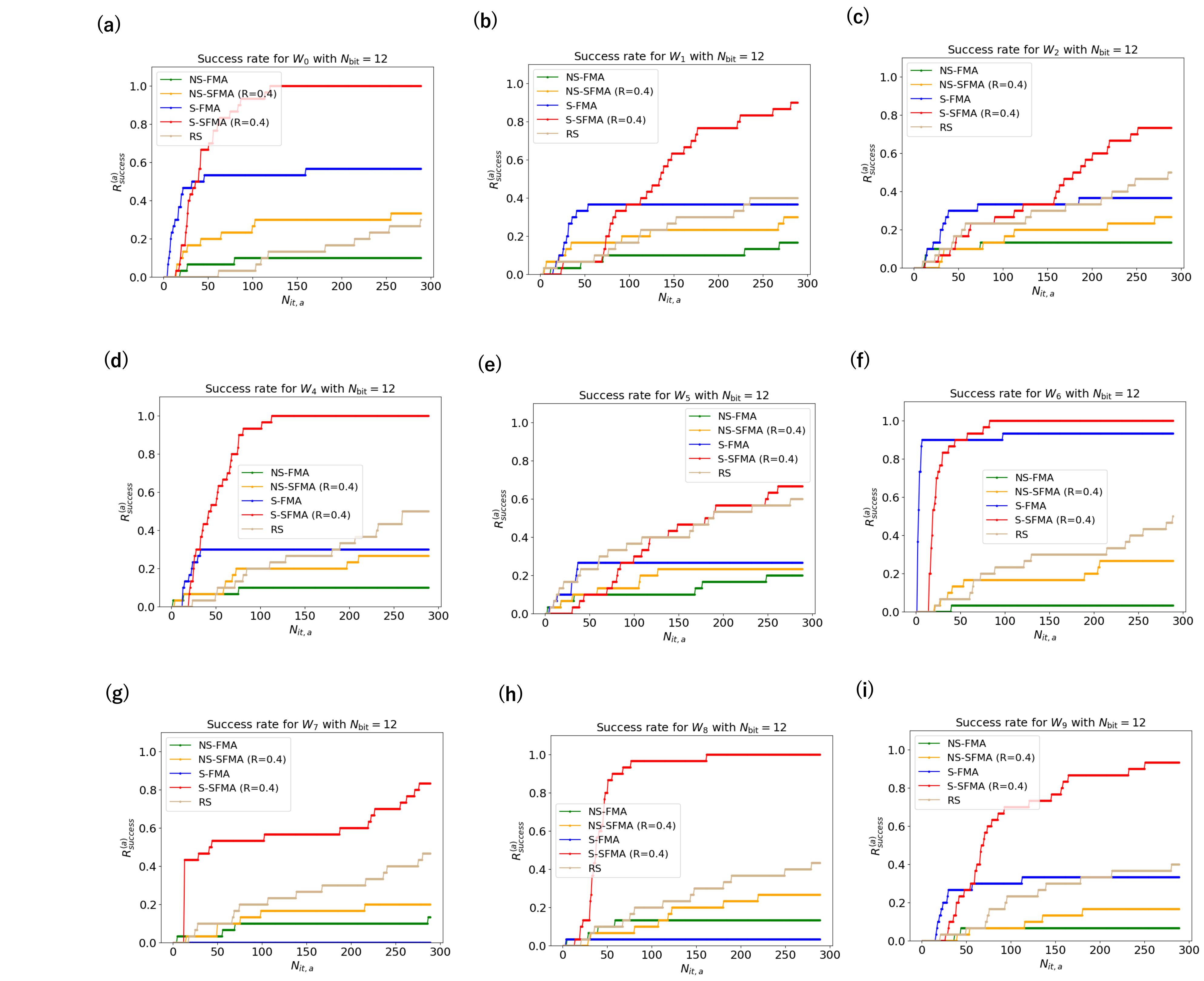}
\end{center}
\caption{Results for  $R_\text{min,success}^{(a)} $ with $N_\text{bit} =12$ and $N_\text{ite}=2N_\text{bit}^2+1$. Plots in  panels 
 (a)-(i)  are the ones for $W_n$ with $n=0,1,2,4,5,6,7,8,$and $9$, respectively.  }
\label{otherresultsWsspin12successrate}
\end{figure*}
\begin{figure*}
\begin{center}
\includegraphics[width=17cm]{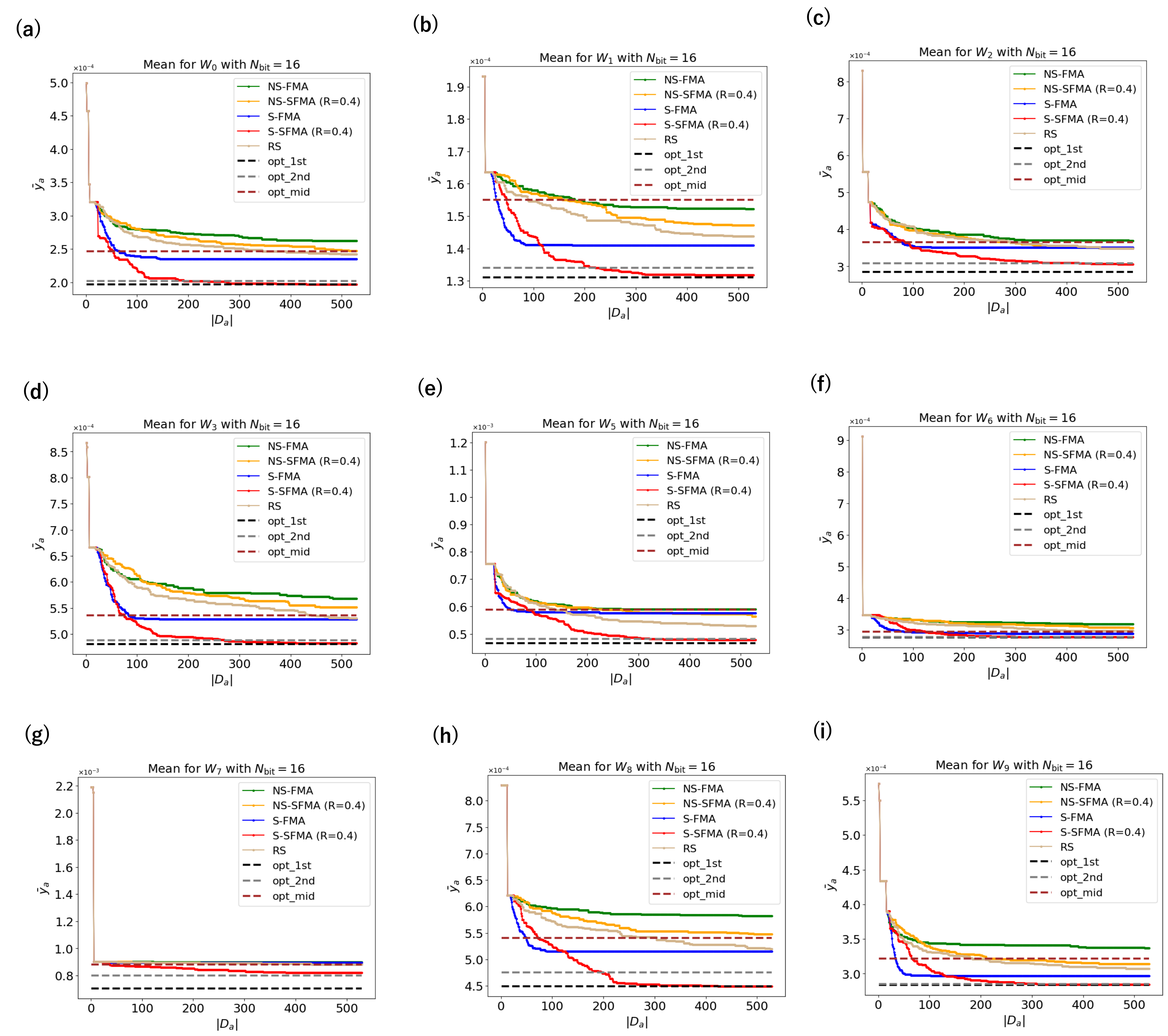}
\end{center}
\caption{Results for  $\bar{y}^{(a)} $ with $N_\text{bit} =16$ and $N_\text{ite}=2N_\text{bit}^2+1$. Plots in  panels
 (a)-(i)  are the ones for $W_n$ with $n=0,1,2,3,5,6,7,8,$ and $9$, respectively.  }
\label{otherresultsWsspin16mean}
\end{figure*}
\begin{figure*}
\begin{center}
\includegraphics[width=17cm]{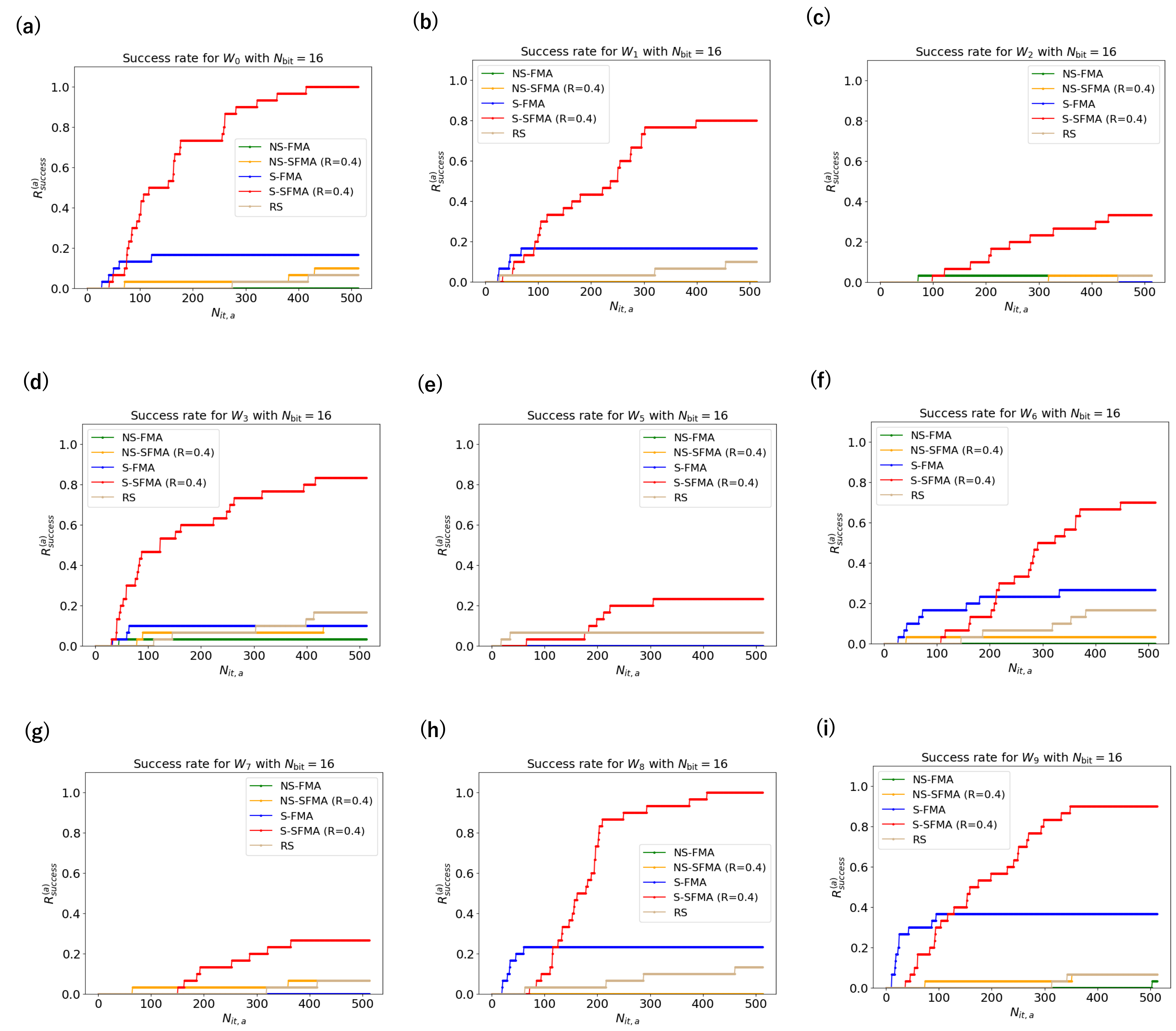}
\end{center}
\caption{Results for  $R_\text{min,success}^{(a)} $ with $N_\text{bit} =16$ and $N_\text{ite}=2N_\text{bit}^2+1$. Plots in  panels
 (a)-(i) are the ones for  $W_n$ with $n=0,1,2,3,5,6,7,8,$ and $9$, respectively.  }
\label{otherresultsWsspin16successrate}
\end{figure*}
\begin{figure*}
\begin{center}
\includegraphics[width=17cm]{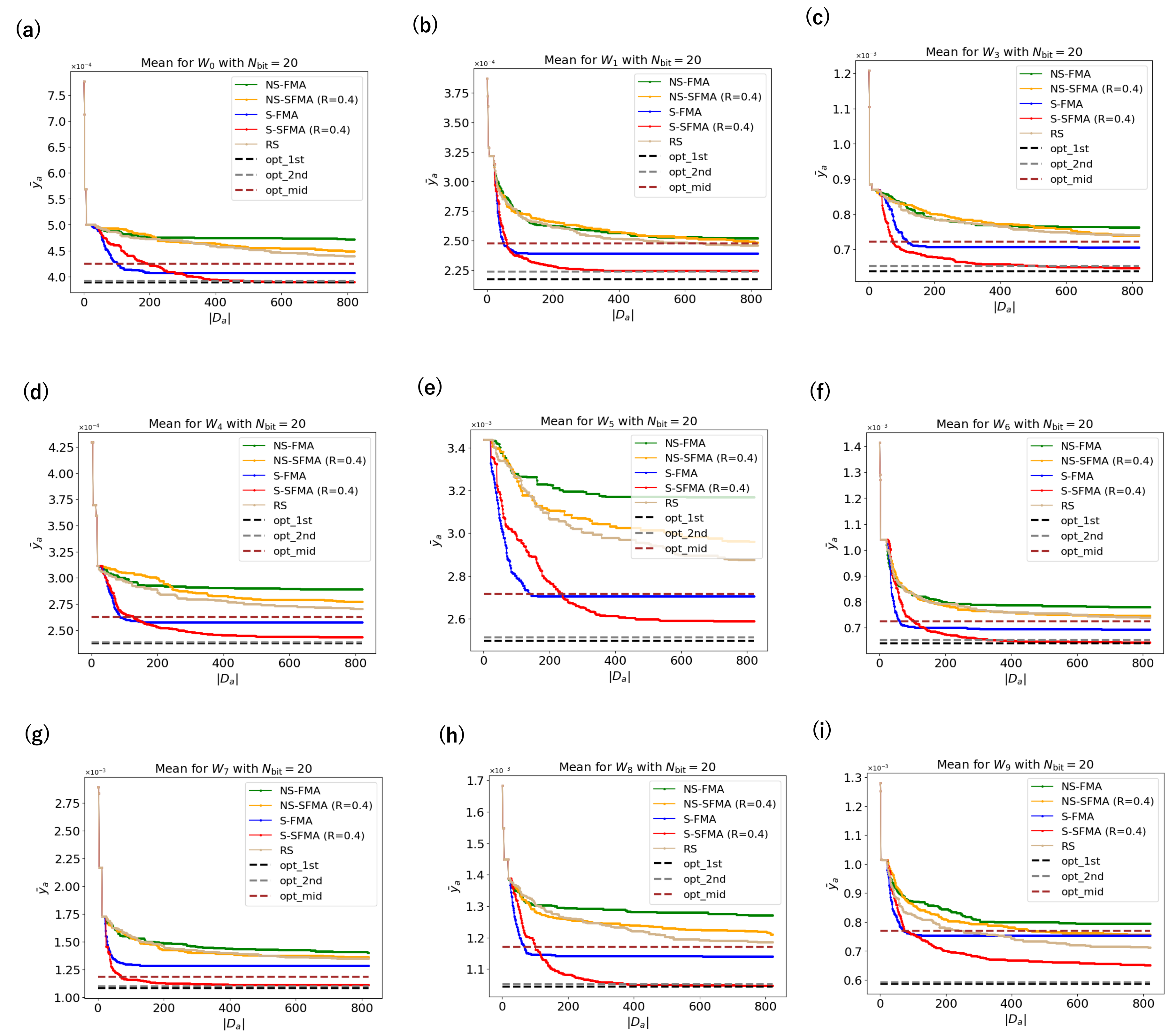}
\end{center}
\caption{Results for  $\bar{y}^{(a)} $ with $N_\text{bit} =20$ and $N_\text{ite}=2N_\text{bit}^2+1$. Plots in  panels
 (a)-(i) are the ones for $W_n$ with $n=0,1,3,4,5,6,7,8$ and $9$, respectively.  }
\label{otherresultsWsspin20mean}
\end{figure*}
\begin{figure*}
\begin{center}
\includegraphics[width=17cm]{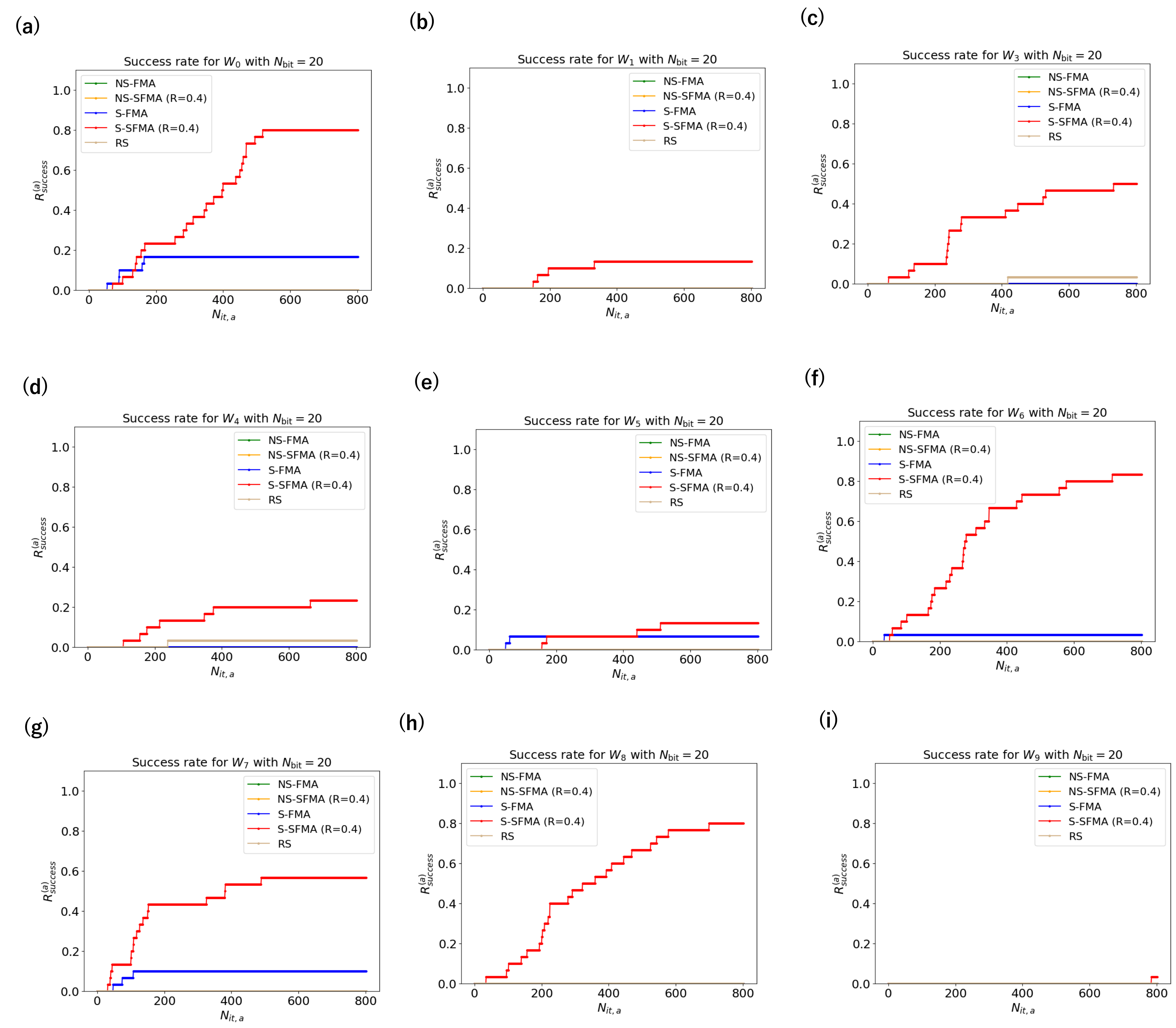}
\end{center}
\caption{Results for  $R_\text{min,success}^{(a)} $ with $N_\text{bit} =20$ and $N_\text{ite}=2N_\text{bit}^2+1$. Plots in  panels
 (a)-(i) are the ones for $W_n$ with $n=0,1,3,4,5,6,7,8$ and $9$, respectively.}
\label{otherresultsWsspin20successrate}
\end{figure*}
  \section{\label{otherresults} Results For The Other $W$ Matrices}
In this Appendix, we explain the results for $\bar{y}^{(a)} $ and $R_\text{min,success}^{(a)} $
for the other nine $W$ matrices that have not been presented in Sec. \ref{numresults}. The plots in Figs. \ref{otherresultsWsspin12mean}, \ref{otherresultsWsspin16mean}, and \ref{otherresultsWsspin20mean} (Figs. \ref{otherresultsWsspin12successrate}, \ref{otherresultsWsspin16successrate}, and \ref{otherresultsWsspin20successrate}) are the results for  $\bar{y}^{(a)} $ ($R_\text{min,success}^{(a)} $) with $N_\text{bit} =12,16,$ and $20$ and $N_\text{ite}=2N_\text{bit}^2+1$, respectively. 
According to the results presented in  Figs. \ref{otherresultsWsspin12mean}-\ref{otherresultsWsspin20successrate} and those in Sec. \ref{numresults}, overall SFMA  outperforms the other algorithms in convergence (SFMA demonstrates the smallest values of $N_\text{conv}$) across the ten problem instances.
Moreover, SFMA demonstrates the advantages in accuracy (SFMA exhibits the largest values of $R_\text{success}^\text{final}$) in  ten  instances for every $N_\text{bit}$.   
From these results, we consider that the advantage of SFMA becomes more prominent as $N_\text{bit}$ increases, particularly, in accuracy.  
Furthermore, RS becomes less effective when $N_\text{bit} $ increases. 

\section{\label{improvedSFMA}  Improved SFMA For Other $W$ Matrices}
In this Appendix, we discuss the improvements of SFMA for other $W$ matrices and present the results for $\bar{y}^{(a)} $ and $R_\text{min,success}^{(a)} $ in Figs. \ref{extendedresultsmeanandratespin12}-\ref{extendedresultsratespin20}.
Figures \ref{extendedresultsmeanandratespin12} and \ref{extendedresultsmeanandratespin16} display the results for  $\bar{y}^{(a)} $ and $R_\text{min,success}^{(a)} $ with $N_\text{bit} =12$ and $16,$ respectively. Meanwhile,  Figs. \ref{extendedresultsmeanspin20} and \ref{extendedresultsratespin20} present those for  $\bar{y}^{(a)} $ and $R_\text{min,success}^{(a)} $ with $N_\text{bit} =20$, respectively.   
Note that,  as explained in Sec. \ref{numresults}, the four algorithms (S-SFMA, S-FMA, NS-SFMA, and NS-FMA) and RS are run such that the datasets with the sizes equal to $2N_\text{bit}^2+1$  are incorporated with  additional data points generated by running the BBO loops
  $2N_\text{bit}^2$ ($4N_\text{bit}^2$) times  for $N_\text{bit}=12$ and 16 ($N_\text{bit}=20$).

Let us begin from explaining the results in Fig. \ref{extendedresultsmeanandratespin12}. Here, we have made the improvements for instances $W_5$ and $W_7$  by taking $R=0.1, N_\text{ite} = 4N_\text{bit}^2+1$.  
 The results demonstrate that the fastest convergence or the lowest value of  $N_\text{conv}$ is achieved by  ISFMA (S-SFMA with $R=0.4$) for $W_5$ ($W_7$). Similarly,  ISFMA (S-SFMA with $R=0.4$) exhibits the highest accuracy for $W_5$ ($W_7$).  
 Next, let us explain the results for $N_\text{bit} =16$ displayed in Fig. \ref{extendedresultsmeanandratespin16}. 
 We have investigated two instances $W_2$ and $W_5$, and for both cases,  ISFMA are created by setting  $R=0.1, N_\text{ite} = 4N_\text{bit}^2+1$. 
As a result, ISFMA only yields  $N_\text{conv}$  and  demonstrates the highest $R_\text{success}^\text{final} $ in these two instances. 
Finally, we discuss the results for $N_\text{bit} =20$. We analyze five cases,  $W_n$ with $n=1,3,4,5,$ and 7, and take $N_\text{ite}$ to be   $N_\text{ite} = 6N_\text{bit}^2+1$.
For $W_n$ with $n=3,5,$ and $7$, we improve SFMA by  taking $R=0.1$, while for the rest we refine SFMA using two different values of $R$   as we have done for $W_9$: see Sec. \ref{numresults}. Namely, for $W_1$ ($W_4$) in the first part we run SFMA with $R=0.1, N_\text{ite}=5N_\text{bit}^2$ ($R=0.1, N_\text{ite}=4N_\text{bit}^2$),  whereas in the second part we run SFMA by tuning the value of $R$ to be $R=0.01$.
As we have done in Sec. \ref{numresults},  we call the improved algorithms given solely by $R=0.1$ ISFMA$_1$ while we call those created by both 
$R=0.1, 0.01$ ISFMA$_2$. Note that ISFMA$_1$ is performed by augmenting the dataset obtained in the first part of the BBO loops in ISFMA$_2$ with the datasets generated by the second parts of the BBO loops. We benchmark the performance of the five (six) algorithms and RS for $W_3, W_5,$ and $W_7$ ($W_1$ and $W_4$). 
As a result,  ISFMA or ISFMA$_i$ ($i=1,2$) do not yield the lowest value of  $N_\text{conv}$ in all the instances; however,  they enable us to achieve the highest values of $R_\text{success}^\text{final}$ in all the instances.      
 Consequently,  we consider that the advantage of the improved SFMA (ISFMA and ISFMA$_i$) in accuracy becomes more prominent when the system size $N_\text{bit}$ gets larger.  
 Moreover,    ISFMA as well as  ISFMA$_i$  outperform the standardized SFMA with $R=0.4$ in  all the  instances   for $N_\text{bit} =16,20$.     %%

\clearpage
\begin{figure*}
\begin{center}
\includegraphics[width=17cm]{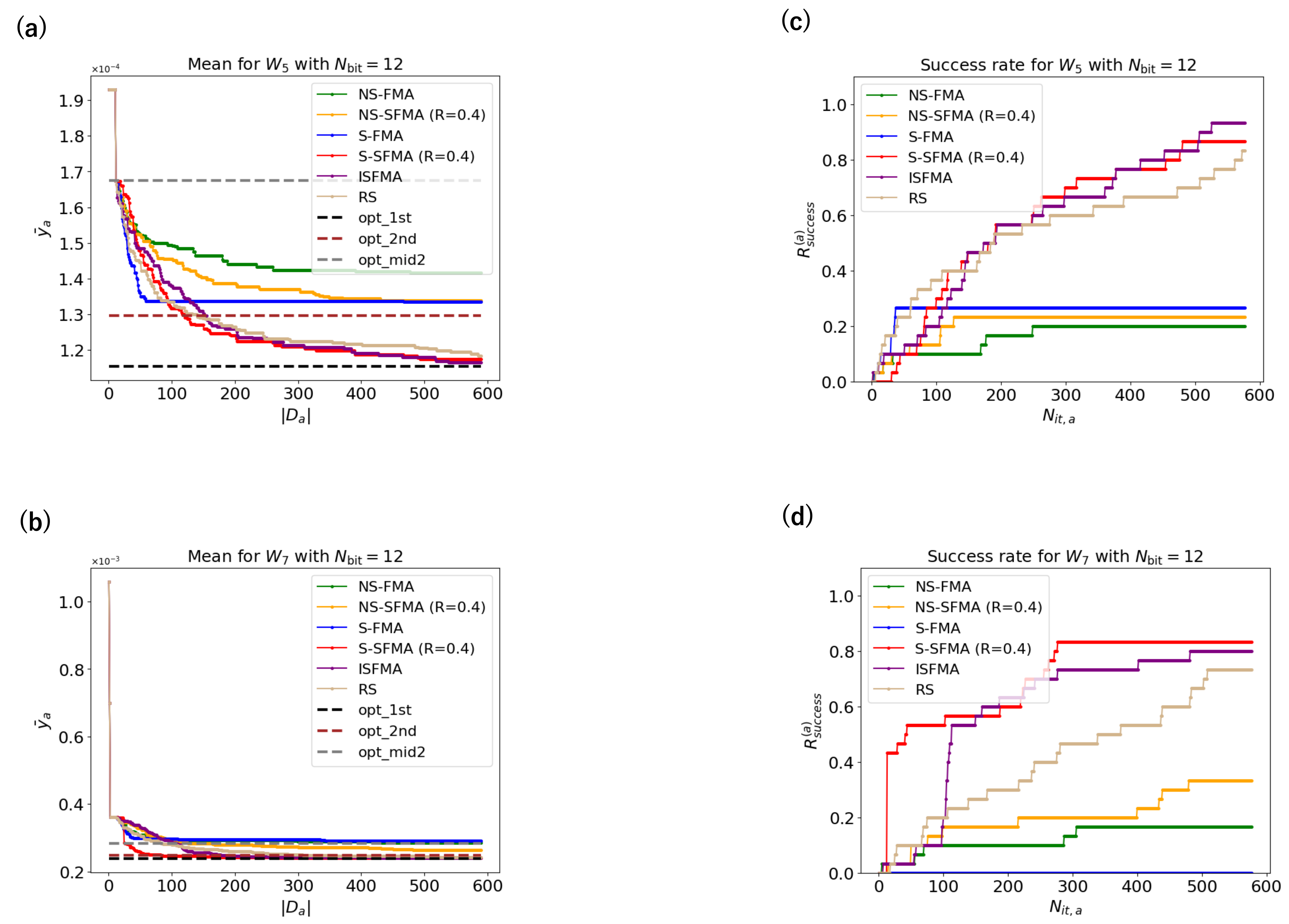}
\end{center}
\caption{Results for $\bar{y}^{(a)} $ and $R_\text{min,success}^{(a)} $ with $N_\text{bit} =12$  and $N_\text{ite}=4N_\text{bit}^2+1$ for
$W_5$ [panels (a) and (c)]  and  $W_7$ [panels (b) and (d)], respectively. The plots in panels (a) and (b) [panels (c) and (d)] represent the results for $\bar{y}^{(a)} $ ($R_\text{min,success}^{(a)}$). }
\label{extendedresultsmeanandratespin12}
\end{figure*}
\begin{figure*}
\begin{center}
\includegraphics[width=17cm]{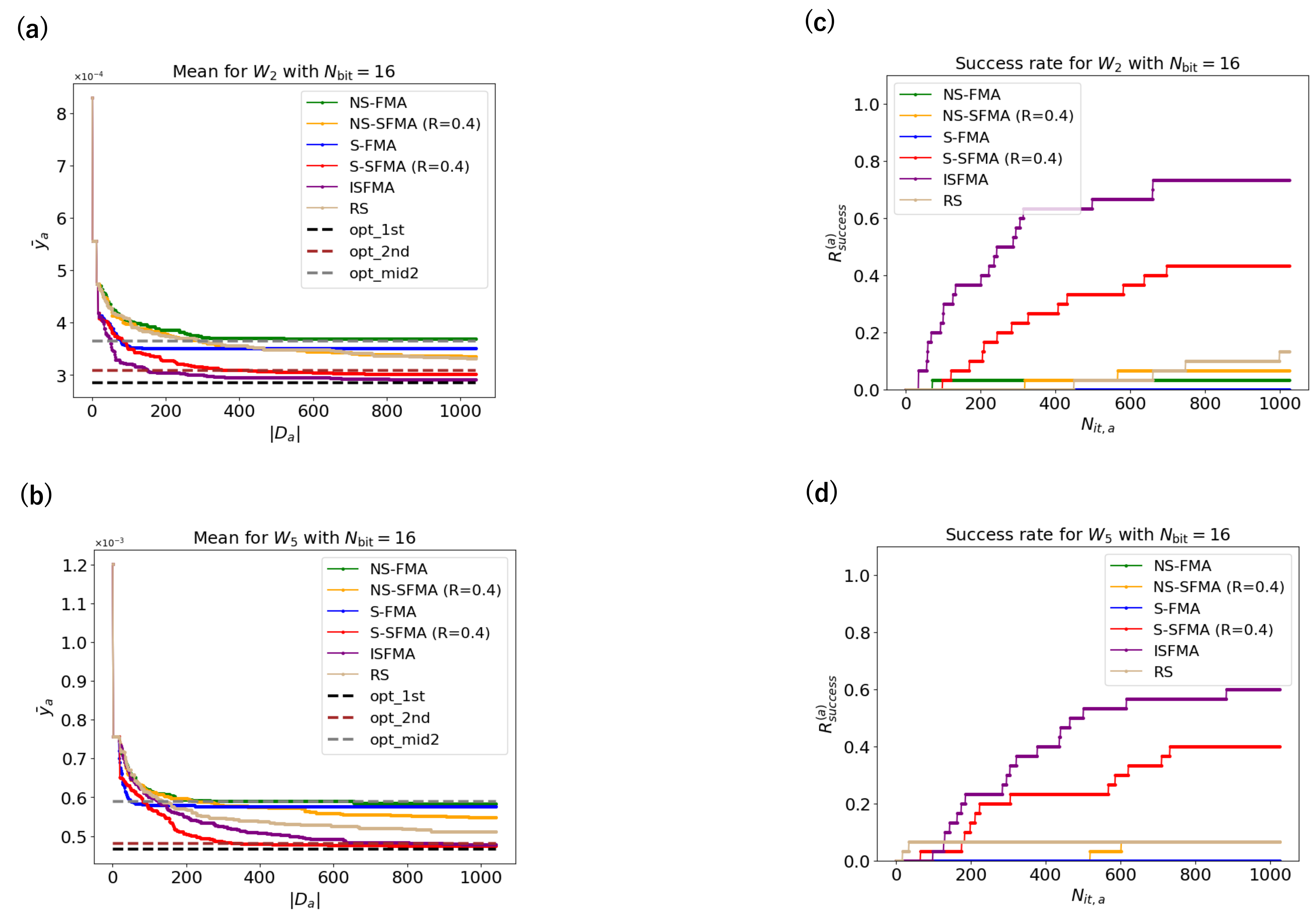}
\end{center}
\caption{Results for  $\bar{y}^{(a)} $ and $R_\text{min,success}^{(a)} $ with $N_\text{bit} =16$  and $N_\text{ite}=4N_\text{bit}^2+1$ for
$W_2$ [panels (a) and (c)]  and  $W_5$ [panels (b) and (d)], respectively. The plots in panels (a) and (b) [panels (c) and (d)] represent the results for $\bar{y}^{(a)} $ ($R_\text{min,success}^{(a)}$). }
\label{extendedresultsmeanandratespin16}
\end{figure*}
\begin{figure*}
\begin{center}
\includegraphics[width=16cm]{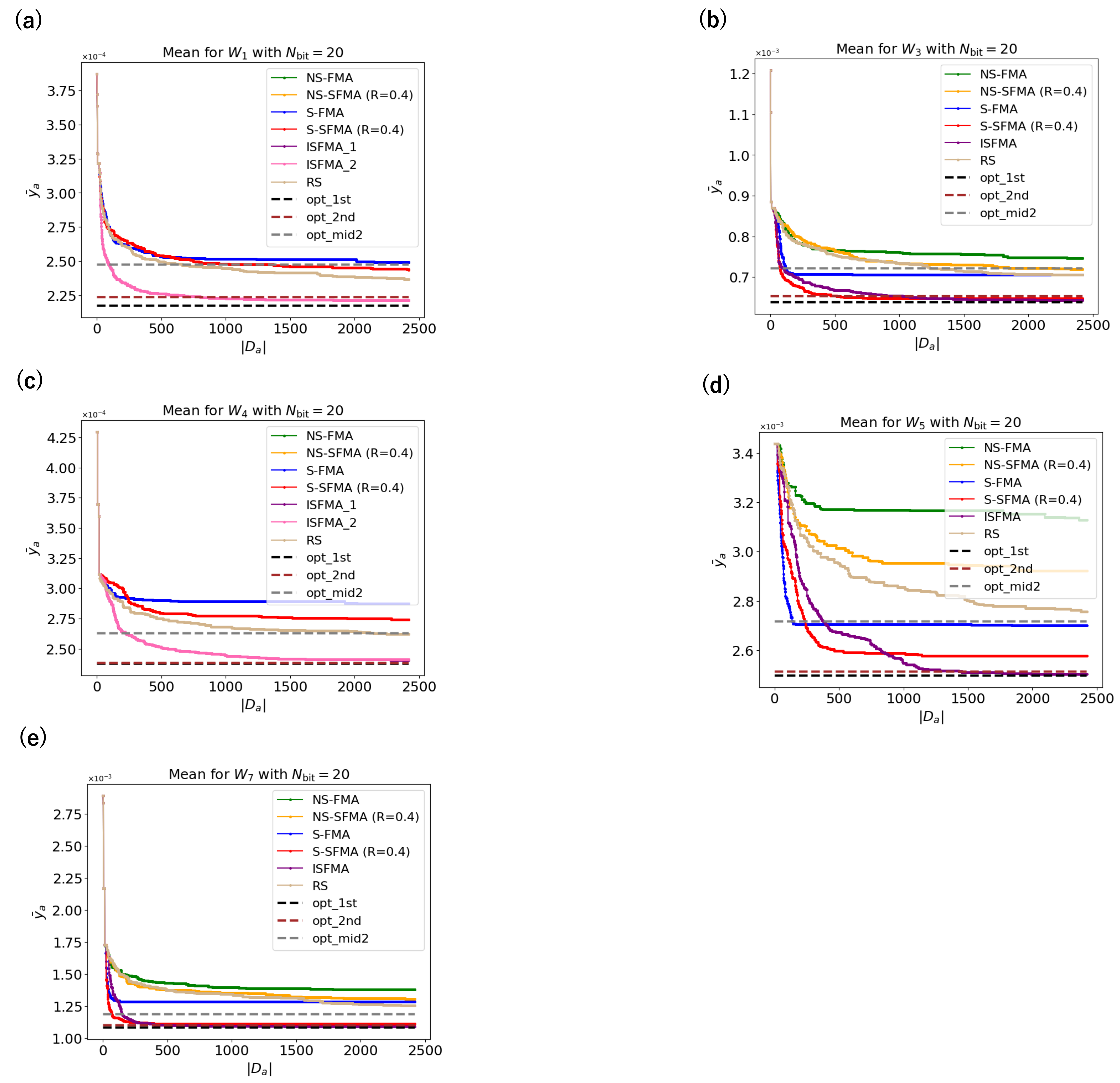}
\end{center}
\caption{Results for  $\bar{y}^{(a)} $ with $N_\text{bit} =20$  and $N_\text{ite}=6N_\text{bit}^2+1$ for
$W_1$ [panel (a)],  $W_3$ [panel (b)], $W_4$ [panel (c)], $W_5$ [panel (d)],  and $W_7$ [panel (e)], respectively.  }
\label{extendedresultsmeanspin20}
\end{figure*}
\begin{figure*}
\begin{center}
\includegraphics[width=16cm]{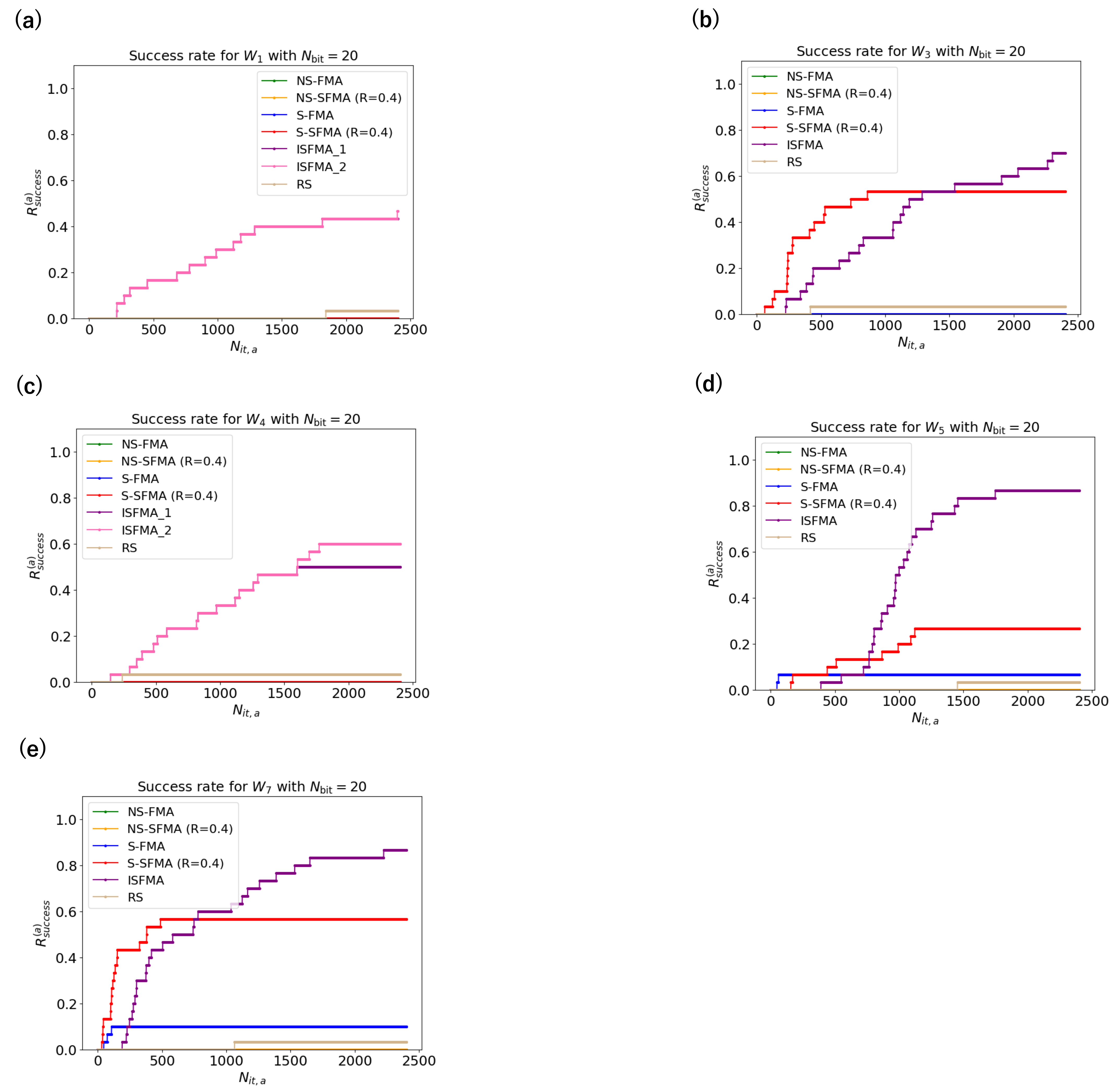}
\end{center}
\caption{Results for  $R_\text{min,success}^{(a)} $ with $N_\text{bit} =20$  and $N_\text{ite}=6N_\text{bit}^2+1$ for
$W_1$ [panel (a)],  $W_3$ [panel (b)], $W_4$ [panel (c)], $W_5$ [panel (d)],  and $W_7$ [panel (e)], respectively.  }
\label{extendedresultsratespin20}
\end{figure*}

\clearpage

\bibliographystyle{unsrtnat} 
\bibliography{RefsSFMAfnl}

\end{document}